\documentclass[12pt, a4paper]{article}

\usepackage{amsmath}
\usepackage{amsfonts}
\usepackage{amssymb}
\usepackage{graphicx, rotating}
\usepackage{epsfig}
\usepackage{latexsym}
\usepackage{graphicx}
\usepackage{color}
\usepackage{amsmath,bm,amssymb}
\usepackage{cite}
\usepackage{slashed}
\usepackage{hyperref}
\hypersetup{colorlinks, citecolor=bluscuro, linkcolor=black, urlcolor=bluscuro}
\definecolor{rossos}{cmyk}{0,1,1,0.55}
\definecolor{bluscuro}{rgb}{0.15, 0.2, .85}
\definecolor{bluchiaro}{cmyk}{1,.3,0.,0.1}

\newcommand{\eq}[1]{Eq.~(\ref{#1})}

\newcommand{\nn}{\nonumber}

\newcommand{\be}{\begin{equation}}
\newcommand{\ee}{\end{equation}}
\newcommand{\bea}{\begin{eqnarray}}
\newcommand{\eea}{\end{eqnarray}}

 \def\ra {\rightarrow}

\newcommand{\arXiv}[2]{\href{http://arxiv.org/pdf/#1}{{\tt [#2/#1]}}}
\newcommand{\arXivold}[1]{\href{http://arxiv.org/pdf/#1}{{\tt [#1]}}}

 \def\la{\lambda}

\begin{document}
\allowdisplaybreaks
\begin{titlepage}
\begin{flushright}
DESY 16-144\\
CERN-TH-2016-170
\end{flushright}
\vspace{.3in}

\vspace{1cm}
\begin{center}
{\Large\bf\color{black}
Interplay of Infrared Divergences and \\[0.5cm] Gauge-Dependence 
 of the Effective Potential}\\
\bigskip\color{black}
\vspace{1cm}{
{\large J.R.~Espinosa$^{a,b}$, M.~Garny$^c$, T.~Konstandin$^d$}
\vspace{0.3cm}
} \\[7mm]
{\it {$^a$\, Institut de F\'isica d'Altes Energies (IFAE), The Barcelona Institute of Science and Technology (BIST), Campus UAB, E-08193, Bellaterra (Barcelona), Spain}}\\
{\it $^b$ {ICREA, Pg. Llu\'is Companys 23, 08010 Barcelona, Spain}}\\
{\it {$^c$\, CERN Theory Department, CH-1211 Geneva, Switzerland}}\\
{\it $^d$ {DESY, Notkestr. 85, 22607 Hamburg, Germany}}
\end{center}
\bigskip

\vspace{.4cm}

\begin{abstract}
The perturbative effective potential suffers infrared (IR) divergences in gauges with massless Goldstones in their minima (like Landau or Fermi gauges) but the problem can be fixed by a suitable resummation of the Goldstone propagators. When the potential minimum is generated radiatively, gauge-independence of the potential at the minimum also requires resummation and we demonstrate that the resummation that solves the IR problem also cures the gauge-dependence issue, showing this explicitly in the Abelian Higgs model in Fermi gauge. In the process we find an IR divergence (in the location of the minimum) specific to Fermi gauge and not appreciated in recent literature. We show that physical observables can still be computed in this gauge and we further show how to get rid of this divergence by a field redefinition. All these results generalize to the Standard Model case.
\end{abstract}
\bigskip

\end{titlepage}

\section{Introduction \label{sec:intro}} 

With the discovery of the Standard Model Higgs during the first LHC run~\cite{higgsD}, it quickly became clear that the precisely measured Higgs \cite{mhLHC} and top \cite{mtcomb} masses point to the possibility of a (very long-lived) metastable electroweak (EW) vacuum \cite{stab0,stab1,stab2,stab3,stab4,stab5}. This fact has triggered a renewed interest on studies (and implications) of the possible metastability of the Standard Model EW vacuum (see e.g. \cite{cosmostab,cosmostab2,deep}).

The main tool for the study of this metastability is the perturbative effective potential \cite{Jackiw}, widely used for studies of spontaneous symmetry breaking. While the effective potential is an enormously useful
tool in such studies, it (or the effective action from which it is derived) is not a physical observable and is subject to gauge-dependence~\cite{Jackiw}. This is a well known issue that has been studied extensively in the literature (see \cite{LitVgauge} for an incomplete list) and is of no serious concern: as long as one is calculating a physical observable (for example the lifetime of the EW vacuum or other tunneling transitions, see e.g. \cite{cosmostab2,Metaxas,IRS,PRM,GK,AFS,Lifetime}), the final answer should be gauge independent.\footnote{In fact, this requirement can be useful to check one is not missing some relevant effect.}
However, this is not always straightforward to achieve in a concrete calculation: usually one must resort to truncations of the perturbative expansion and this can jeopardize the gauge independence of the final result. 

A well known example of this
kind of problem occurs in the Coleman-Weinberg model of radiative symmetry breaking \cite{CW} (as we review in Subsection \ref{sec:CW}) or in the Standard Model (SM), as the instability that appears in the potential at high field values is generated radiatively. In order to have the gauge dependence of the potential under control in such cases one must resort to resummations of series of corrections to the potential of arbitrarily high order, as nicely demonstrated in \cite{AFS}. This type of resummation is reminiscent of the  resummation of the Goldstone propagator needed to solve 
the infrared (IR) problem of the effective potential due to Goldstone contributions \cite{Martin,EEK} in those gauges that feature massless Goldstone bosons at the potential minima. One of the main results of this paper is that the resummation required to fix the IR problem automatically takes care of the resummation needed to control the gauge-dependence issues (Subsection \ref{sec:CW}).

To check explicitly the gauge-independence of the observables derived from the effective action one must resort to families of gauges. The most common gauge choices are the $R_\xi$ and Fermi gauges, both of which contain a gauge-fixing parameter (or parameters, that we will generically call $\xi$) that can be used to keep track of the gauge-dependence. In this paper we follow a large fraction of recent literature and use Fermi gauge for this purpose. In order to keep the analysis transparent, we work in the Abelian Higgs model (Section~\ref{sec:Abelian}). The results we obtain can be extended to the SM in a straightforward way
 (Section \ref{sec:SM}).

In our analysis we find that Fermi gauge is afflicted by an IR divergence\footnote{Apparently this has escaped the attention of recent literature but was already known, see e.g. \cite{Loinaz}.
In the context of the Nielsen identity, the IR troubles with Fermi gauge  were remarked even earlier \cite{Nielsen,Aitchison}.} which is absent in $R_\xi$ or Landau gauges (and can be traced back to the mixing of the Goldstone bosons with the gauge bosons). More specifically, the first derivative of the effective potential is logarithmically divergent for vanishing Goldstone mass (as happens in the broken minimum). Naively, this  is a 
severe problem, since the minimum of the potential determines the vacuum of the system and is found by solving $\partial V/\partial h=0$. Furthermore, we show that this divergence persists even if the Goldstone propagator is resummed~(section \ref{sec:IRresum}). However, we also show that observables like the Higgs mass are  IR finite (section \ref{sec:MH}), and likewise the Nielsen identity still holds (section \ref{sec:Nielsen}). We take this good behaviour as an indication that Fermi gauge is not sick and we then present a way to obtain an explicitly 
IR-finite effective potential by a suitable  rescaling of the Higgs field (section \ref{sec:shift}). We draw some general conclusions in Section \ref{sec:concl} and leave some more technical details to a few Appendices.

\section{IR Problems and Gauge Dependence 
 \label{sec:IRpro}}

Before discussing the gauge
dependence of the effective potential we  address the
infrared problems associated with the presence of massless Goldstone modes at the potential minima. This problem was recognized in \cite{OtherIR},
which identified IR divergent contributions to the effective potential from loops involving Goldstone bosons. The solution to
this problem, in Landau gauge, is simply to resum the Goldstone
contributions by the appropriate shift of the Goldstone two-point function as was first proposed in \cite{Martin,EEK} (see \cite{Pila,Martin2,Kim} for later developments and applications). 
This simple resummation makes the Landau-gauge potential and its first derivative IR finite.  As an added bonus, it turns out (see Subsection~\ref{sec:CW}) that this resummation not only fixes the IR problems of the effective potential but it also resolves the issues with residual gauge-dependence in those potentials that feature extremal points generated radiatively.

In what follows we want to apply this resummation prescription to the effective potential in  Fermi gauge, starting with the Abelian Higgs Model and generalizing later on to the SM case. As we show explicitly below, the resummed potential is IR finite as desired. However, the first derivative, unlike
what happened in Landau gauge, is still IR divergent. We 
also provide a solution to this problem in Section~\ref{sec:shift}.

\subsection{Abelian Higgs Model in Fermi Gauge\label{sec:Abelian}} 

For simplicity, let us start the discussion of infrared problems in the Abelian Higgs model. The Lagrangian, in Fermi gauge, reads
\be
  {\cal L}=-\frac14 F_{\mu\nu}F^{\mu\nu}-\frac{1}{2\xi}(\partial^\mu B_\mu)^2 + |D_\mu\phi|^2 + m^2\phi^\dag\phi - \la (\phi^\dag\phi)^2 \,,
\label{LagAbel}
\ee
where the covariant derivative for the charged ``Higgs'' field 
\be
  \phi = \frac{1}{\sqrt{2}}\left(h + i \chi \right)\,,
\ee 
is 
\be
  D_\mu = \partial_\mu - i\,gB_\mu\, .
\ee
Without loss of generality we take the charge of $\phi$ under the $U(1)$ gauge symmetry to be unity and $F_{\mu\nu}$ is the corresponding field strength.
The Lagrangian in Eq.~(\ref{LagAbel}) includes the
gauge-fixing term, which corresponds to the so-called Fermi (or Lorentz) gauge, and depends on a free parameter, $\xi$. The limit $\xi\to 0$ corresponds to Landau gauge.

The one-loop effective potential for this Abelian Higgs model was
first derived long ago by Dolan and Jackiw, in their classic paper \cite{DJ}. Its explicit expression requires 
knowing the masses in a generic field background $h$, in
which one has:\footnote{To simplify later expressions, here and in the following Sections we use capital letters, as defined above, to denote field-dependent squared masses.}
\bea
\label{masses}
  G&\equiv & m_\chi^2 = -m^2 + \la h^2 \,,\nn\\ 
  H &\equiv &m_h^2 = -m^2 + 3\la h^2 \,,\\ 
  B &\equiv &m_B^2 = g^2 h^2 \,.\nn
\eea

Using dimensional regularization, with $d=4-2\epsilon$, 
the one-loop effective potential is then given by
\be
  V_1 = i \int \frac{d^dk}{(2\pi)^d} \Bigg( \sum_{\mbox{\tiny fermions}\atop\mbox{\tiny ghosts}} \log \det i{\cal G}^{-1} - \frac12 \sum_{\mbox{\tiny bosons}} \log \det i{\cal G}^{-1} \Bigg)\ ,
\ee
where $i{\cal G}^{-1}$ denote the inverse of the propagators.
For the Abelian Higgs model in Fermi gauge in $d$ dimensions, one has contributions from transverse gauge bosons, Higgs and mixed Goldstone-longitudinal gauge bosons, giving:
\bea
  \log \det i {\cal G}^{-1}_T &=& (d-1)\log(-k^2+B) \,,\nn\\
  \log \det i {\cal G}^{-1}_h &=& \log(-k^2+H) \,,\nn\\
  \log \det i {\cal G}^{-1}_L &=& \log[-k^2(k^4-k^2 G+\xi B G)] \nn\\
  &=& \log(-k^2+G_+) + \log(-k^2+G_-) + \log(-k^2)\ ,
  \label{Vmom}
\eea
with
\be
  G_\pm = \frac12 \left(G \pm \sqrt{G^2-4\xi B G} \right)\ ,
  \label{Gpm}
\ee
while the ghost-contribution is independent of the field value, $h$.
Performing the momentum integrals, the final form of the renormalized effective potential, in $\overline{\rm MS}$  scheme, is:
\bea
 V_{1} &=& \frac{\kappa}{4}\left[
 3B^2\left(L_B-\frac{5}{6}\right)
 +H^2 \left(L_H-\frac{3}{2}\right)
 +\ G_+^2 \left(L_{G_+}-\frac{3}{2}
 \right)+
 G_-^2 \left(L_{G_-}-\frac{3}{2}\right)\right]\ ,
 \label{VeffAbel}
 \eea
 where $\kappa\equiv 1/(16\pi^2)$, $L_X\equiv \log(X/\mu^2)$ and $\mu$ is the $\overline{MS}$ renormalization scale. This agrees with the expression given in \cite{DiLuzio}, translated to the Abelian case.
The Landau-gauge limit corresponds to $\xi\rightarrow 0$, which
gives $G_+\rightarrow G$ and $G_-\rightarrow 0$.

In analytical expansions below we will consider only the case $\xi>0$,
which is usually better behaved than $\xi<0$ \cite{DiLuzio}. Eq.~(\ref{Gpm}) shows that, for $\xi>0$, $G_\pm$ become imaginary in some field range even when $G>0$. However, the corresponding imaginary parts cancel each other out and  the potential itself stays real.\footnote{This is most clearly seen in the unintegrated expression involving ${\cal G}^{-1}_L$ in 
Eq.~(\ref{Vmom}). After the Wick rotation to Euclidean momentum $k^2\rightarrow -k_E^2$, one has $\log[-k^2(k^4-k^2 G+\xi B G)]\rightarrow \log[k_E^2(k_E^4+k_E^2 G+\xi B G)]$
and the argument of the logarithm is positive for $\xi>0$ (and $G>0$).}

When one approaches the potential vacuum, $h\rightarrow v$ (with $v^2=m^2/\lambda$ at tree-level),
the Goldstone mass goes to zero, $G\rightarrow 0$, and generically this induces IR divergences in the effective potential. In Landau gauge, the potential $V$ first develops IR divergences at 3-loop order \cite{OtherIR}. The trouble comes from potential terms that are schematically of the form $\delta V\sim X^2 \log G$, where $X$ is some mass-squared that is nonzero at the minimum of the potential. On the other hand, the first derivative $V'$ of the potential is IR divergent already at 2-loop order, from terms in the potential of the form  $\delta V\sim XG\log G$. Finally, the second derivative $V''$ is IR divergent already at 1-loop, from a term $\delta V \sim G^2\log G$. 

This infrared behaviour is worse in Fermi gauge than in Landau
gauge.  The troublesome terms are of similar origin but replacing $G$
by $G_\pm$ and one sees that the terms $\delta V\sim G_\pm^2\log G_\pm$ cause $V'$ to diverge already at one loop, due to the fact that $G_\pm'$ diverges and $G_\pm'G_\pm$ goes to a nonzero constant for $G\to 0$. To see this most clearly,
notice that for $h\rightarrow v$ one has $G_\pm \rightarrow \pm i\sqrt{\xi G B}$, so that $G_\pm^2\log G_\pm\sim X G\log G$. More precisely the source of the trouble is the term
\be
\delta V_{1} = -\frac{\kappa}{4}\xi G B \log\left(\frac{\xi G B}{\mu^4}\right)\, .
\ee
In the following Subsection we apply to this Fermi-gauge case the resummation proposed in \cite{Martin, EEK} to cure 
such IR problems. 

Let us close this Subsection with a brief discussion of the gauge dependence of the potential in Eq.~(\ref{VeffAbel}).  In spite of the 
explicit $\xi$-dependence of the potential through its dependence
on $G_\pm$, it is well known that the value of the potential at its extremal points
is guaranteed to be gauge-invariant by the Nielsen identity \cite{Nielsen,FukudaKugo,Aitchison}. At one-loop order this is obviously the case of 
the potential in Eq.~(\ref{VeffAbel}): the only dependence on $\xi$ of the potential appears in the $G_\pm$ terms and at the minimum 
$G\rightarrow 0$ one has $G_\pm \rightarrow 0$, so that the $\xi$ dependence disappears.

\subsection{IR-Resummation \label{sec:IRresum}}

The IR divergences in the effective potential are due to massless Goldstones, $G\rightarrow 0$, and come from diagrams with Goldstone bosons that carry small momentum, $k^2\sim G$. The worst divergences originate from those diagrams that have the largest possible number of Goldstone propagators with the same small momentum, and this number grows with higher loop order. As shown in detail in Refs.~\cite{Martin,EEK} for the SM in Landau gauge, these Goldstone divergences are spurious and can be resummed in a simple way by reorganizing the perturbative expansion. This is done by including the effect of self-energy diagrams on the Goldstone propagators, with $G\rightarrow \overline G\equiv G + \Pi_g$,  where $\Pi_g$ is a well-defined radiative contribution to the Goldstone squared-mass 
that can be calculated perturbatively to the order needed. As explained in \cite{EEK}, $\Pi_g$ includes only contributions from heavy fields (that is, fields whose mass does not vanish when $G\rightarrow 0$) and hard Goldstones (with momentum $k^2\gg G$). 

The effect of resummation in the Goldstone contribution to the one-loop potential in Landau gauge is therefore the replacement\footnote{For our purposes in this paper the one loop resummed result 
(\ref{Vre}) will be enough, but the resummation procedure can be extended to higher orders, see \cite{EEK} for details.
}
\be
\delta_G V = \frac{\kappa}{4}G^2(L_G-3/2)\quad
\to\quad \delta_G \overline V=\frac{\kappa}{4}\overline G^2(L_{\overline G}-3/2)\ .
\label{Vre}
\ee
Expanding the latter expression perturbatively (in powers of $\kappa$) indeed reproduces the IR divergent terms of the unresummed potential. In the unresummed perturbative expansion, the IR divergences occur at the field value for which $G\rightarrow 0$: the location of the tree-level minimum. In the resummed potential, instead, possible IR divergences would occur at $\overline G\rightarrow 0$, which corresponds to the minimum of the radiatively corrected potential. However, for the resummed potential, $\overline V$ and $\overline V'$ are IR finite and only the second derivative $\overline V''$ diverges for $\overline G\rightarrow 0$. However, this divergence is harmless and in fact required to get right the physical Higgs mass, as we discuss below.

The generic resummation of IR divergences just reviewed can also be applied to the Fermi gauge. The small complications associated with gauge boson-Goldstone mixed propagators can be circumvented in a simple way: add and subtract to the Lagrangian a term\footnote{A more sophisticated procedure is described in Subsection \ref{sec:Nielsen}, see in particular footnote~\ref{footchi2}.} $-\Pi_g \chi^2/2$, where $\Pi_g$ is the (zero-momentum) two-point function for the Goldstone field $\chi$ obtained as discussed above. The explicit expression for $\Pi_g$ in the Abelian model at one-loop is
\be
\Pi_g=3\kappa\left[g^2B\left(L_B-\frac{1}{3}\right)
 +\lambda H \left(L_H-1\right)\right]\ ,
 \label{Pig}
\ee
which can be directly obtained from the contribution of 
$B$ and $H$ to the one-loop potential in Eq.~(\ref{VeffAbel})
remembering that the Goldstone mass is given by $(\partial V/\partial h)/h$.
The added term is treated as shifting the Goldstone mass that appears in propagators, with $G\rightarrow \overline G\equiv G+\Pi_g$,  while the subtracted term is treated as a counterterm. 
After this shift, the two field-dependent
masses corresponding to the mixed Goldstone-gauge boson sector
are given by
\be
\overline G_\pm \equiv\frac{1}{2}\left(\overline G\pm\sqrt{\overline G^2-4\xi \overline G B}\right) \ ,
\ee
to be compared with Eq.~(\ref{Gpm}).
The expression for the one-loop effective potential of Eq.~(\ref{VeffAbel}) with this resummation 
implemented is obtained from the unresummed one simply by the replacement $G_\pm\rightarrow \overline G_\pm$.

Does this resummation achieve the desired cure of the IR divergence problems of the effective potential also in Fermi gauge?  While it is clear 
that  the resummed potential is finite in the
$\overline G\rightarrow 0$ limit, its first derivative (which is 
the crucial quantity to determine the location of the potential minimum) is not finite even after resummation. The unresummed
term that causes the IR divergence in $V'$ is of the form $\delta V\sim XG\log G$ as discussed at the end of the previous Section and resummation
simply changes this to $\delta\overline V\sim X\overline G\log \overline G$, which still gives a divergent $\overline V'$.  Note that the divergence in $\overline V'$ is $\xi$-dependent and goes away
for $\xi=0$. This divergence, which can be translated into a divergence in the one-loop vacuum expectation value (vev), had
been pointed out before (see e.g. \cite{Loinaz}) but seems to have gone unnoticed in more recent literature. 
Before discussing the solution to the previous problem (deferred to Subsection~\ref{sec:shift} below), it is instructive to compare the resummation performed above with the resummation discussed in Ref.~\cite{AFS} to solve a different issue.

\subsection{IR Resummation Eliminates Residual Gauge Dependence \label{sec:CW}}

Suppose we are interested in the gauge dependence of the effective potential close to the electroweak vacuum. 
The perturbative counting is the conventional loop counting, with $g^2 \sim \lambda$. The naive expectation is that, using this counting, a consistent expansion of the effective action will fulfill the Nielsen identity~\cite{Nielsen} and hence provide gauge-independent observables. As we have described above, potentially this requires resummation of certain classes of diagrams~\cite{GK,Martin,EEK}, most notably two-particle-reducible diagrams of light particles in those gauges (like Landau or Fermi gauge) in which the Goldstone boson is massless at the minimum. 

Things are different if the electroweak vacuum is generated radiatively. The best known example is the Coleman-Weinberg Model \cite{CW}, which is nothing but 
the Abelian Higgs model with a massless scalar $h$ [that is, $m^2=0$ in the Lagrangian of Eq.~(\ref{LagAbel})]. Famously, the interest of the model lies in the possibility of radiatively breaking  the $U(1)$ gauge symmetry (a paradigmatic example of dimensional transmutation).
For studies of the gauge dependence of the effective potential, 
the difficulty with this model was recognized already in Refs.~\cite{Nielsen,FukudaKugo}: the minimum appears through the balance between the tree-level quartic coupling $\lambda$ and 
the one-loop radiative corrections, of order $\hbar g^4$,
so that for power counting one should use $\lambda \sim
\hbar g^4$. This jeopardizes the usual fixed-order loop expansion of the effective potential: one-loop terms of order $\hbar \lambda g^2$ are of the same order as two-loop terms of
order $\hbar^2 g^6$ or three-loop terms of order $\hbar^3 g^{10}/\lambda$, and this should be taken into account when 
showing the gauge independence of the value of the potential 
at its  minimum,  that would have a residual gauge dependence if calculated at a fixed order in perturbation theory. It was also clear \cite{Johnston} that a resummation  that reorganizes the perturbative expansion would get rid of this problem and this has been shown explicitly
to two-loop order in Ref.~\cite{AFS}.

A similar situation arises in the SM effective potential  at very high values of the Higgs field, when an instability is generated by radiative corrections. Previous work in Fermi gauge has studied the gauge dependence of the potential at such high field values \cite{DiLuzio,AFS}, at which one can neglect the explicit mass term in the Lagrangian (that is of electroweak scale size) and use the counting $\lambda \simeq \hbar g^4 \simeq \hbar y_t^4$. This simplifies the analysis of the effective action, since the one-loop corrections to the effective potential from the Goldstone bosons scale as $\lambda^2$ and hence are of the same order as three-loop contributions from the gauge sector $\sim g^8$. Thus, in a two-loop analysis up to order $g^6 \sim \lambda^{3/2}$ some of the IR issues do not enter yet. 
But even in this simplified case, the same subtleties discussed for the Coleman-Weinberg model concerning gauge-dependence remain, as emphasized in \cite{AFS}.

We now show that the resummation required to cure the IR
problems of the potential discussed in the previous Subsection  automatically takes care of this gauge issue. To ease the comparison, note that the notation of \cite{AFS} for the Coleman-Weinberg Model in Fermi gauge, differs from ours: $h$ is called $\phi$, $g$ is $e$ and our $\lambda$ is replaced by $\lambda/6$. The resummation shift in the Goldstone mass, 
$G=\lambda h^2 \rightarrow \overline G=G+\Pi_g$, corresponds to the shift $\lambda\rightarrow \bar\lambda(h)\equiv \lambda - \hat\lambda(h)$, where 
\be
\hat\lambda(h)\equiv 36\kappa g^4\left(1-3L_B\right)\ ,
\ee 
with $\kappa=\hbar/(16\pi^2)$ and $L_B=\log(g^2h^2/\mu^2)$. We have used here the results of the previous Section setting $m^2=0$ and neglecting $\lambda^2$ corrections in Eq.~(\ref{Pig}), which are of higher order as $\lambda \sim\hbar g^4$.

The expression for the resummed two-loop potential is quite simple:
\bea
\overline V_{2}(h) &=& \frac{1}{4}\lambda h^4\nonumber\\
&+&\frac{3\kappa}{4} g^4h^4\left(L_B-\frac{5}{6}\right) -\frac{\kappa}{4}\xi
g^2\bar\lambda(h) h^4\left(\log\frac{\xi g^2\bar\lambda(h) h^4}{\mu^4}-3\right)\nonumber\\
&+&\kappa^2 g^6h^4 \left(\frac52 L_B^2
-\frac{31}{3}L_B+\frac{71}{6}\right)\ .
\label{V2r}
\eea
Some comments are in order:
{\it (1)}  This expression packages in a compact way the result for the 2-loop potential given in \cite{AFS}, Eqs.~(6.16-17).
{\it (2)} The resummation performed to deal with IR divergences generates directly all the two-loop terms
necessary to check gauge independence, without residual gauge dependence left.
{\it (3)} The gauge independence of the potential value at the minimum is  straightforward to see as the result of the minimum corresponding (at one-loop order) to $\bar\lambda(v)=0$.

The resummed expression given in Eq.~(\ref{V2r}) also sheds some light
on the IR problem in $V'$ discussed in the previous Subsection as specific to Fermi gauge.
Indeed there is a ($\xi$-dependent) logarithmic divergence in
$dV/dh\sim \kappa\xi g^2 (d\bar\lambda/dh)\log\bar\lambda$
for $\bar\lambda\rightarrow 0$. Moreover, notice 
that $d\bar\lambda/dh=-d\hat\lambda/dh\sim \kappa g^4$, so that the IR divergence is a two-loop effect of order $\kappa g^6$
and no other terms of that order in the potential (\ref{V2r}) could cancel out such divergence.

Nevertheless, this obstruction looks strange, given the fact that the first derivative of the potential $V$, to whatever precision  it is calculated, and the Goldstone mass calculated to the same precision, are related\footnote{As the effective potential in a generic background is a function of $|\phi|^2=(h^2+\chi^2)/2$.} as $G=(dV/dh)/h$. In a 
consistent calculation there seems to be no room for a zero in
$G$ causing a divergence in $dV/dh$. In view of this, we could consider \eq{V2r} as the correct expression for the 2-loop resummed potential but with $\bar\lambda$ to be specified in a self-consistent way. Then we could use the relation
$\overline G =\bar\lambda(h)h^2 = (\partial \overline V_{2}/\partial h)/h$
to define $\bar\lambda$. At two loops, with the approximations used, one gets
\bea
\bar\lambda(h)&=&\lambda +\kappa g^4\left(3L_B-1\right)+
2\kappa \xi \lambda g^2\left(1-L_B\right)
\nonumber\\
&+& \kappa^2 g^6\left\{\frac23\left[40-\left(47-15L_B\right)L_B\right]
+\xi\left[1+\left(5-6L_B\right)L_B\right]
\right\}\nonumber\\
&-&\kappa\xi g^2\left[\lambda+\frac12\kappa g^4\left(1+6L_B\right)\right]
\log(\xi\bar\lambda/g^2)\ .
\label{barlambda}
\eea
However, this definition of $\bar\lambda(h)$ is problematic for $\bar\lambda\rightarrow 0$ as the prefactor of the last logarithm does not go to zero in that limit.
Another way of stating the problem is this: fix the values of $\lambda$, $g$ and $\xi$ at some given $\mu$. Using \eq{barlambda} as the definition of the function $\bar\lambda(h)$, we see that such function cannot cross zero, implying there is no extremum in $V(h)$. When $\bar\lambda$ gets close to zero, $\log\bar\lambda$ blows up and destroys perturbativity. 

The ultimate root of this IR divergence is the pole of order $p^4$ in the mixed propagator of Goldstone bosons and longitudinal gauge bosons, which shows up clearly in their contribution to the effective potential, see Eq.~(\ref{Vmom}).\footnote{A similar $p^4$ pole appears in supersymmetric QED (in the propagator of the lowest component of the vector superfield), leading to IR divergences. For a recent discussion see \cite{Dine}.}
Moreover,  this problem persists even if no perturbative expansion is used (as we show in Appendix~\ref{App:Ward} using
the Ward identity). This mixed propagator is a specific feature of  Fermi gauge which explains why  such IR divergence is absent in Landau gauge ($\xi \to 0$) or in the background $R_\xi$ gauges.

Naively one might be tempted to conclude that there is no acceptable description of symmetry breaking within perturbation theory in Fermi gauge (unless $\xi=0$, which is Landau gauge).
Nevertheless, as physical quantities cannot depend on the gauge parameter $\xi$, one could expect that the $\xi$-dependent IR divergence will cancel out when calculating observables. In the following Section we show that this expectation is fulfilled for the physical Higgs mass.

\section{Physical Results in Fermi Gauge}

In this Section we show how physical information can be extracted in Fermi gauge even though the effective potential has no well-defined minimum at one-loop order. First we discuss the mass of the Higgs boson and then how to make sense of the Nielsen identity in spite of the IR divergences that afflict Fermi gauge.

\subsection{The Physical Higgs Mass \label{sec:MH}} 

We show now how the physical Higgs mass is free of IR divergences
even when one calculates it in the Fermi gauge.
We go back here to the general Abelian Higgs Model, with 
nonzero~$m^2$.  
The physical Higgs mass is defined as the pole of the Higgs propagator.  Calculated at one loop order it is
\be
M_h^2 = V_{0}''(v) + \Sigma(M_h^2)|_0 
=-m^2 + 3\lambda v_0^2 + 6 \lambda v_0 \delta v + \Sigma(M_h^2)|_0 \, .
\label{Mhpole}
\ee
Here,  field derivatives are represented by primes; the one-loop vev is $v=v_0+\delta v$ with $v_0=\sqrt{m^2/\lambda}$ the tree-level vacuum expectation value [calculated with the tree-level potential $V_0(h)$] and $\delta v$ the one-loop correction to it, given by
\be
\delta v = -\frac{1}{2\lambda v_0^2}\,\left.\frac{\partial V_{1}}{\partial h}\right|_0 \, .
\ee
Finally, $\Sigma(p^2)$ is the one-loop 1PI two-point function of the Higgs, with external momentum $p$.
With $|_0$ we indicate that the limit $h\to v_0$ is taken, which is appropriate at one-loop order. All the parameters entering Eq.~(\ref{Mhpole}) are already the renormalized ones. 

The explicit result in Fermi gauge, using dimensional regularization, $L_X\equiv \log(X/\mu^2)$, and taking the limit $h\to v_0$ everywhere except in the logarithmically
divergent terms, yields (see Appendix \ref{App:Higgs} for details)
\be
  6 \lambda v_0 \delta v|_0 = 3\kappa v_0^2\left\{
   6\lambda^2(1-L_H)+g^4(1-3L_B) 
  \left.{} -\frac{1}{2}\lambda g^2\xi\left[2-\log\left(\frac{\xi G B}{\mu^4}\right)\right]\right\}\right|_0 \, , 
  \label{vshift}
\ee
and
\bea
 \Sigma(M_h^2)|_0 &=& \kappa v_0^2\left\{
 3g^4(1+L_B)+\lambda g^2\left[2+3\xi-2L_B-\frac32\xi \log\left(\frac{\xi G B}{\mu^4}\right)\right] \right.\nn\\
 &&\left.\left.{}+
 6\lambda^2\left(\pi\sqrt{3}-7+4L_H\right) 
 - 2(\lambda^2-2\lambda g^2+3g^4) B^R_0(B,B,H)
\frac{}{}\right\}\right|_0 \, .
\label{SigmaH}
\eea
where we have left unevaluated the (renormalized) one-loop integral
\be\label{eq:oneloopintegral}
B^R_0(m_1^2,m_2^2,p^2) = 
-
 \int_0^1 dx \log\left[\frac{m_1^2(1-x)+m_2^2 x-x(1-x)p^2-i\varepsilon}{\mu^2}\right]\ ,
\ee
and we leave explicit the terms that cause a divergence in the $G\ra 0$ limit.
In the sum that gives $M_h^2$, the terms involving $\xi$ in (\ref{vshift}) and (\ref{SigmaH}) cancel, 
as expected for a physical quantity \cite{Gambino}. Furthermore, one can check that the result above for  $M_h^2$ agrees with the SM result calculated in Landau gauge\footnote{
Starting from Eqs.~(2.12-13) of \cite{stab0}, one gets $M_h^2 = 2\lambda v_0^2 + 4 \lambda v_0 \delta v^{(\xi=0)} + \delta_1 M_h^2$, where $\delta v^{(\xi=0)}$ is the shift of the minimum in Landau gauge. Taking into account the different conventions used,  and translating to the Abelian case ($y_t\to 0, g\to 0, {g'}^2\to g^2/4$, $M_Z^2\to B$, and ignoring terms involving $M_W$) there is agreement. }\cite{stab0,CEQR} and Feynman gauge \cite{SZ}, appropriately reduced to the Abelian Higgs model.

Note that the terms that diverge logarithmically for $h\to v_0$ (i.e. for $G\to 0$), that is,  the IR divergence from the shift in the Higgs vev (\ref{vshift}) and from the self-energy (\ref{SigmaH}), cancel in the sum, and one has
\be
-3\frac{V_{1}'}{v_0} +\Sigma(M_h^2) \Rightarrow {\rm IR\ finite}.
\label{VpS}
\ee
Therefore, even though 
the Higgs vev (not an observable) diverges, observable physical quantities as the Higgs mass are 
finite.\footnote{This cancellation is reminiscent of a similar cancellation that takes
place in the computation of the Higgs mass as $M_h^2 = V''+\Sigma(M_h^2)-\Sigma(0)$ where $V$ is now the full potential  and the $\Sigma$ terms take into account
that the mass is defined on-shell and not at $p^2=0$.
At one loop, $V_{1}''$ has a logarithmic IR divergence that 
is precisely cancelled by $\Sigma_1(M_h^2)-\Sigma_1(0)$,
as it is obvious from the fact that $V_{1}''$ is nothing 
but $\Sigma_1(0)$.} 
 We also note that the imaginary part of the pole of the Higgs propagator (\ref{Mhpole}), that is related to the Higgs width, is gauge independent. Specifically, after the cancellation of the terms involving $\xi$ in (\ref{vshift}) and (\ref{SigmaH}), the only contribution to the imaginary part arises from
the last term in (\ref{SigmaH}) that involves the one-loop integral (\ref{eq:oneloopintegral}). This term is independent of $\xi$, and has a non-zero imaginary part
for $H>4B$ (that is, $m_H>2m_B$), corresponding to the decay of the Higgs into a pair of gauge bosons.

We close this Subsection with am illustrative comparison of the IR divergences in the 1PI two-point function $\Sigma(p^2)$ between Landau and Fermi gauges. In Landau gauge the IR divergent terms at one-loop are
\bea
\kappa^{-1}\Sigma_{IR}(p^2)&=&0\ , \quad\quad\quad\quad ({\rm for}\ p^2\neq 0)\nn\\
\kappa^{-1}\Sigma_{IR}(0)&=& 2 \lambda^2 v^2 L_G\ . 
\eea
This divergent structure is correlated with the fact that, in this gauge,  $V'$ is IR finite and $V''$ IR divergent,
as $V_{1}''=\Sigma(0)$.

In Fermi gauge, instead, we have
\bea
\kappa^{-1}\Sigma_{IR}(p^2)&=& \frac{\xi B L_G}{2v^2}\left[\frac{1}{p^2}(H-p^2)^2\left(1+\frac{\xi B}{p^2}\right)-
5H+2p^2\right]\ , \quad ({\rm for}\ p^2\neq 0)\nn\\
\kappa^{-1}\Sigma_{IR}(0)&=& - \lambda^2 v^2 \xi \frac{B}{G}-
\frac{3\pi}{4}\lambda^2 v^2\sqrt{\frac{\xi B}{G}}+
\left(\lambda^2 v^2-\frac{5}{2}\lambda \xi B\right)
L_G\ . 
\label{IRSigma}
\eea
We see that $\Sigma$ is IR divergent even on-shell, and this is correlated (in order to get an IR finite Higgs mass as discussed above) with the IR divergence in  $V'$ present in this gauge, see Eq.~(\ref{VpS}).
We also see that the $\xi\to 0$ limit of $\Sigma_{IR}(0)$ differs from the Landau gauge result showing explicitly that the limits $\xi\to 0$ and $p^2 \to 0$ do not commute.

\subsection{Nielsen Identity\label{sec:Nielsen}}

The Nielsen identity \cite{Nielsen,FukudaKugo,Aitchison} plays a central role for the gauge (in)dependence of the potential. 
In this Subsection we examine how the IR divergence in $V'$ affects this identity, which reads
\be
  \xi \frac{\partial V}{\partial\xi}+C\frac{\partial V}{\partial h} = 0\,,
  \label{NI}
\ee
where the function $C$ is the constant background limit of a function $C(x)$ (which enters the Nielsen identity for the effective action) that in Fermi gauge reads
\be\label{eq:NielsenC}
  C(x) = \frac{i g}{2} \int  d^4 y \, \langle c(x) \chi(x) \bar c(y) \partial_\mu B^\mu(y) \rangle\, ,
\ee
with $c, \bar c$ the ghost fields.
The (renormalized) one-loop result reads
\be\label{eq:NielsenC1}
  C_1 =
   \frac{\kappa \xi B}{2v(G_+-G_-)}\left[G_+(L_{G_+}-1)-
   G_-(L_{G_-}-1)\right]\ .
\ee
Taking the limit $h\to v$ (or $G\to 0$), we find
\be 
 C_1\to -\frac12 \kappa \xi g^2v\left[ 1 - \frac12\log\left(\frac{\xi G B}{\mu^4}\right) \right]
 \ ,
 \label{CIR}
\ee
which is logarithmically IR divergent for $G\to 0$.\footnote{This was noticed already in \cite{Nielsen} and was later 
taken as reason to avoid the use of Fermi gauge, {\it e.g.} in \cite{Aitchison}.} However, evaluating the Nielsen identity perturbatively, at one-loop one gets (with primes denoting field derivatives):
\be
\xi\frac{\partial V_{1}}{\partial \xi} + C_1  V'_{0}=0\ ,
\label{NI1}
\ee
and  the one-loop product $C_1 V_{0}' \propto C_1 G$ goes to zero for
$G\to 0$. This means that the value of $V$ at the minimum, or more precisely $V_{1}|_{v}$, is gauge independent, as it should be.

The fact that the (one-loop) Nielsen identity (valid for arbitrary field values) is IR finite implies that
all the identities derived from it by taking field derivatives are
also IR finite even if individual terms diverge. For example,
taking one field derivative of the Nielsen identity  gives
\be
  \xi \frac{\partial V'}{\partial\xi}+C V''+ C' V' = 0\, ,
\ee
When evaluated close to the potential minimum, $h=v$, the first term essentially determines the gauge dependence
of the location of that minimum. One can then check that the IR divergences in the first two terms cancel each other.
Naively, one may think that the last term vanishes at $v$. However, this is not the case because $C'\propto 1/G$
such that the product $C'V'_{0}$ does not vanish in the minimum.

A similar discussion applies to the Nielsen identity for the kinetic term $Z(h) (\partial_\mu h)^2/2$ in the effective action (derived
in Appendix \ref{app:NielsenKin} for Fermi gauge at one-loop). As in Landau gauge, $Z(h)$ is IR divergent close to the broken phase minimum. The enhanced IR sensitivity of $Z$ can be attributed to the gradient expansion around homogeneous field configurations as well as the vanishing Goldstone boson mass in the broken minimum, that occurs both in Landau and Fermi gauges. Nevertheless, the IR sensitivity of the coefficients appearing in the Nielsen identity for $Z$ matches precisely those that are present in $\xi\partial Z/\partial\xi$,
and the Nielsen identity holds at one-loop for all field values. In addition, as discussed above, all IR divergences cancel for physical observables.

We conclude that the Nielsen identity is fulfilled order by order in the perturbative expansion
and that,  for the effective potential, IR effects explicitly cancel.
Finally, it can be shown that the identity will also hold after including resummation effects in a consistent way. For instance, the one-loop resummed potential 
fulfills the identity as the replacement $G\ra \overline G=G+\Pi_g$ (or $G_\pm\ra \overline G_\pm$), with a $\xi$-independent $\Pi_g$, does not interfere with the structure of the identity if one uses, to be consistent, that $V'=h \overline G$. In fact, there is a systematic way to consistently maintain the Nielsen identity order by order even when including IR resummation in the following way: Add to the Lagrangian $-\Pi_g \phi^\dag\phi +\Pi_g \phi^\dag\phi=0$, and absorb the first term into a shift of the quadratic term $-m^2\to -m^2+\Pi_g$ that enters in the Goldstone and Higgs mass. On the other hand, the second term is treated perturbatively as a one-loop counterterm, i.e. ${\cal O(\hbar)}$. This procedure implements the replacement $G\to \overline G$ in the one-loop expressions and, in addition, also the Higgs mass gets shifted. This ensures that the potential at order $\hbar^0$ fulfills $V'=h \overline G$, and therefore the Nielsen identity is
precisely fulfilled.\footnote{\label{footchi2}Note that for this to be true it is not sufficient to add and subtract only a mass term for the Goldstone boson $-(\Pi_g-\Pi_g)\chi^2/2$. The reason is that this
operator is not gauge invariant, and therefore jeopardizes the Nielsen identity once the first term is resummed while the second is treated as a one-loop counterterm. Using instead the operator $\phi^\dag\phi$ solves this problem.} On the other hand, since $\phi^\dag\phi=(h^2+\chi^2)/2$ this also introduces a shift in the Higgs mass parameter $H\to H+\Pi_g$ that appears in the Higgs propagator in loop diagrams. However, this shift is perturbatively small (since the Higgs mass does not vanish close to the broken phase minimum) and therefore it is cancelled by the corresponding counterterm contributions up to terms of higher order in perturbation theory. 

\section{Solutions to the IR Problem in Fermi Gauge \label{sec:shift}}

Although, as we have seen, IR divergences should not affect observables, it can be convenient to get rid of the IR divergences
also in intermediate results. In particular, it is more
satisfactory to have an effective potential whose first derivative is IR finite and does not suffer from an infinite shift in the location of its minimum. With such goal in mind, an obvious solution to try is to absorb the infinite shift in a field redefinition.

\subsection{Field Redefinition}
 
Let us see then what field redefinition would be needed to make $V'$ IR finite. Consider a field redefinition of the form
\be
h \to h + \hbar F(h)\ ,
\label{shift}
\ee
where we have included an explicit factor $\hbar$ to indicate that the shift is of one-loop order. Such field redefinitions modify
the form of the Lagrangian without affecting the physics (more precisely, without modifying $S$-matrix elements \cite{shift}).
The change induced by the shift in Eq.~(\ref{shift}) on the potential is:
\be
  V = V_{0}  + \hbar V_{1}  +{\cal O}(\hbar^2)\to V = V_{0}  + V_{0}'\hbar F + \hbar 
  V_{1}  +{\cal O}(\hbar^2)\ ,
\label{shiftV}
\ee
with primes denoting field derivatives, as usual. There is some freedom in choosing the function $F(h)$ so that it cancels out the IR divergence in $V_{1}'$, because the cancellation should occur at a single point in field space. 
The IR divergence comes from the potential term $\delta V_{1}=-(\kappa/4)\xi G B \log (\xi G B/\mu^4)$ and we have the choice of removing this, or just the part that goes like $\log(G/\mu^2)$ or the full Goldstone contribution, etc. We do not commit at this point with such choices and write generically the term to be removed as $\delta V_{1}=-(\kappa/4)\xi G B \log (G/X)$ with $X$, a quantity of dimension mass squared, to be chosen later on.
Simple inspection of the first derivative of the shifted potential in Eq.~(\ref{shiftV}) shows that $F(h)$ evaluated at $h=v$ should satisfy 
\be
F(v) = \left.\frac{\kappa}{4}\xi g^2 v\log \frac{G}{X}\right|_v\ .
\label{cond1}
\ee

Once we get an IR finite $V_{1}'$, it can no longer
cancel the IR divergence of $\Sigma(M_h^2)$ as needed to get $M_h^2$ finite, see Eq.~(\ref{VpS}) and the discussion at the end of Subsection \ref{sec:MH}. However, this causes no problem as we should also consider the impact of the field shift (\ref{shift}) on the kinetic term for $h$: 
\be
\frac12 (\partial_\mu h)^2 \to \frac12 (\partial_\mu h)^2 +
\hbar  F'  (\partial_\mu h)^2\ ,
\label{shiftK}
\ee
that modifies the Higgs pole mass equation (\ref{Mhpole}) as
\be
(1+2F') M_h^2 = [V_{0} + F V_{0}']''(v)   + \Sigma(M_h^2)|_0\ .
\label{Mhpoles}
\ee
In this equation we have treated the $F$-shift terms  in (\ref{shiftV}) and (\ref{shiftK}) as modifying the tree-level Lagrangian (even though they are shifts of order $\hbar$). The one-loop radiative corrections calculated with this shifted Lagrangian are the same as before so that $\Sigma(M_h^2)|_0$ above is the same as in  (\ref{Mhpole}),
and the one-loop vev $v$ is the minimum of the full one-loop potential (\ref{shiftV}). Explicitly,
\be
v=v_0+\delta_F v + \delta v\ ,
\ee
with 
\be
v_0^2=\frac{m^2}{\lambda}
\ ,\quad\quad
\delta_F v = -F
\ ,\quad\quad
\delta v = -\frac{V_{1}'}{V_{0}''}
\ ,
\ee
so that $\delta v$ is the same as in (\ref{Mhpole}) but now $\delta_F v + \delta v$ is IR finite by construction. More explicitly,
expanding $v$ in Eq.~(\ref{Mhpoles}) and keeping terms up to ${\cal O}(\hbar)$ we get
\be
(1+2F') M_h^2 = V''_{0}(v_0) +(\delta_F v +\delta v)  V'''_{0}+ 2 F' V''_{0} + F V_{0}'''   + \Sigma(M_h^2)|_0\ .
\label{Mhpoles2}
\ee 
The $F'$ terms cancel out and we end up with
\be
M_h^2 = 2\lambda v_0^2 +3\lambda v_0 (\delta_F v +\delta v) +  [3\lambda v_0 F + \Sigma(M_h^2)|_0]\ .
\ee
We see explicitly that $M_h^2$ is exactly the same as the one in (\ref{Mhpole}) as $3\lambda v_0 (\delta_F v + F) =0$. Moreover,
the IR divergence of $\Sigma(M_h^2)|_0$ is precisely cancelled by the $3\lambda v_0 F $ term. Therefore, we see explicitly that the physical Higgs mass, an observable, is not affected by our field redefinition, as expected.

A simple and convenient choice of $F(h)$ that satisfies the condition (\ref{cond1}) corresponds to the field redefinition
\be
h \to h + \frac{\kappa}{4}\xi g^2 h\log \frac{\xi G B}{\mu^4}\ ,
\label{hred}
\ee
corresponding to the choice $X=\mu^4/(\xi B)$.
This field redefinition, being $\mu$-dependent, modifies the
wave-function renormalization of the field, encoded in the anomalous dimension $\gamma \equiv d\log h/d\log\mu$. One gets
that the one-loop $\gamma$ is shifted as:
\be
\gamma_1 =\kappa g^2(3-\xi) \to \gamma_1=\kappa g^2(3-\xi)  +\kappa \xi g^2 = 3 \kappa g^2\ ,
\ee
and the $\xi$ dependence drops (which can be useful to reduce the gauge dependence of the potential). However, as we have already mentioned, one is not forced to this choice of $X$ and if one takes instead a $\mu$ independent $X$ (say $X=B$) the one-loop
anomalous dimension of $h$ will not change.

The potential expressed in terms of the shifted field 
[after $h\to h + (\kappa/4)\xi g^2 h\log(G/X)$] reads,
at one-loop:
\bea
V &=& -\frac12 m^2h^2 + \frac{\lambda}{4}h^4 +
\frac{\kappa}{4}\left[
 3B^2\left(\log\frac{B}{\mu^2}-\frac{5}{6}\right)
 +H^2 \left(\log\frac{H}{\mu^2}-\frac{3}{2}\right)\right]
\nn\\
&+&
 \frac{\kappa}{8}G\left[G\left(\log\frac{\xi G B}{\mu^4}-3\right)-2\xi B\left(\log\frac{\xi X B}{\mu^4}-3\right) + (G_+-G_-)\log\frac{G_+}{G_-}\right]\ .
\eea
One can explicitly check that $V'$ is now finite for $G\to 0$ and the one-loop shift of the vev is IR finite. The field redefinition we have performed is reminiscent of the field redefinition proposed in \cite{NielsenRedef} to obtain a $\xi$-independent
potential (which in practice is equivalent to going to Landau gauge)
or in \cite{cosmostab2} to make the field canonical and reduce the $\xi$ dependence of the potential. Our aim here is different: we just want to remove the IR problem but leave the $\xi$ dependence as we want to study the gauge (in)dependence of different quantities. 
Still, one could argue that our field redefinitions either fix the gauge (if all $\xi$ dependence is gone) or amount to using a different gauge fixing (in some sense intermediate between Fermi and Landau gauges).

Let us consider next the IR-structure of the Nielsen identity after
the field redefinition in (\ref{hred}). At one-loop, the identity
takes the form in Eq.~(\ref{NI1}).
As the field shift sends $V_{1}\to V_{1} + F V'_{0}$ we immediately deduce that $C_1\to C_1-\xi\partial F/\partial \xi$:
\be
C_1 \to C_1 -\frac{\kappa}{4} \xi g^2 h \left(\log\frac{G}{X}+1\right)\ ,
\ee
so that the Nielsen identity is respected.
From the IR limit in Eq.~(\ref{CIR}) we see that the new $C_1$
above is instead IR finite. Thus, the same field redefinition that removes the IR divergence from the first derivative of the effective potential also removes the IR divergence in the Nielsen coefficient.
In addition, when taking a derivative of the shifted Nielsen identity with respect to the shifted field, the contribution $C' V_{0}'$ actually vanishes for $h \to v_0$, in accordance with naive expectations.

\subsection{IR Regulator}

Above we have shown how to use a field redefinition to get a potential with a well behaved (i.e. IR finite) first derivative. Instead, we could as well simply regulate the IR divergences, checking at the end of the calculations that physical quantities are independent of the IR regulator. A simple way of doing this 
is to use the Fukuda-Kugo gauge \cite{FukudaKugo}
\be
 {\cal L}_{gf} = -\frac{1}{2\xi}\left(\partial^\mu B_\mu + \mu_{IR} \chi\right)^2\,,
\ee
that leads to the masses 
\be
G_\pm = \frac12 \left[ G+2\mu_{IR} m_B \pm \sqrt{G^2-4(\xi B-\mu_{IR} m_B) G} \right] \, ,
\ee
which tend to $\mu_{IR} m_B$ when $G\to 0$.
Therefore, in this gauge $\mu_{IR}$ acts as an IR regulator of the divergences
that afflict Fermi gauge (recovered at $\mu_{IR}\to 0$).

\section{The Standard Model in Fermi Gauge \label{sec:SM}}

It is an straightforward exercise to extend the results for
the Abelian Higgs model, discussed in the previous Sections,
to the non-Abelian case and in particular to the SM. 
The (electroweak) gauge-fixing terms in the Lagrangian, in Fermi gauge, are
\be
{\cal L}_{\rm gf}=-\frac{1}{2\xi_B}(\partial^\mu B_\mu)^2 -\frac{1}{2\xi_W}(\partial^\mu W^a_\mu)^2 \ ,
\ee
where $B_\mu$ is now the $U(1)_Y$ gauge boson and 
$W_\mu^a$ are the $SU(2)_L$ ones. The Higgs doublet, with hypercharge $Y=1/2$, is written
as 
\be
H=\left(\begin{matrix}
\chi^+\\
\frac{1}{\sqrt{2}}(h+i\chi)
\end{matrix}
\right)\ .
\ee
The potential is a function of the neutral field $h$
and the $\chi,\chi^\pm$ fields are the three Goldstones.

The renormalized $\overline{MS}$ effective potential, calculated up to one loop order, has the form
\be
V = -\frac12 m^2 h^2 + \frac14 \lambda h^4 +
\frac{\kappa}{4}\sum_\alpha N_\alpha M_\alpha^4(h)
\left(\log\frac{M_\alpha^2(h)}{\mu^2}-C_\alpha\right) \ ,
\ee
where $\alpha$ runs over all particle species, with $N_\alpha$ counting the corresponding degrees of freedom (taken negative for fermions) and tree-level mass-squared $M^2_\alpha(h)$ in the $h$ background.
The $C_\alpha$ are constants (equal to 3/2 for scalars and fermions, and to 5/6 for gauge bosons). The particle species and masses relevant
for the potential are:
\be
\hspace*{-0.5cm}
\begin{array}{rll}
{\rm Top\ quark:} &  N_t=-12\ , & T\equiv M_t^2 = \frac12 y_t^2 h^2\ , \\
W^\pm : & N_W=6\ , & W\equiv M_W^2 = \frac14 g^2 h^2\ , \\
Z^0 : & N_Z=3\ , & Z\equiv M_Z^2 = \frac14 (g^2+{g'}^2) h^2\ , \\
{\rm Higgs} : & N_h=1\ , & H\equiv M_h^2 = -m^2 + 3\lambda h^2\ , \\
{\rm Charged\ Goldstones} : & N_{A_\pm}=2\ , & G_{A_\pm}\equiv M_{A_\pm}^2 = \frac12(G\pm\sqrt{G^2-4\xi_W G W})\ , \\
{\rm Neutral\ Goldstones} : & N_{B_\pm}=1\ , & G_{B_\pm}\equiv M_{B_\pm}^2 = \frac12[G\pm\sqrt{G^2-4(\xi_W W+\xi_B B)G}]\ , 
\end{array}
\ee
where we have used the auxiliary squared masses
\be
B\equiv \frac14 {g'}^2h^2 = Z-W\ ,\quad G\equiv 
-m^2 + \lambda h^2\ .
\ee
As in the Abelian case, the minimum of the tree-level
potential corresponds to $G=0$. The above expression for the effective potential is well known, see e.g. \cite{DiLuzio,AFS}.

The IR properties of the potential in the limit $G\to 0$ are 
similar to those discussed in the Abelian model. There are 
IR divergences at higher orders in the perturbative expansion of the potential that can be resummed as in Landau gauge \cite{Martin,EEK}, by the shift $G\to G+\Pi_g$, where now,
at one-loop order
\be
\Pi_g=3\lambda H(L_H-1)-6y_t^2 T(L_T-1)+\frac32 g^2 W (L_W-1/3)+\frac34 (g^2+{g'}^2)Z(L_Z-1/3)\ ,
\ee
where $L_X=\log(X/\mu^2)$. This resummation also solves the issue of residual gauge dependence at high field values in the region of instability \cite{AFS}, as discussed in Subsection~\ref{sec:CW}.

As in the Abelian case, however, this resummation still leaves
a potential that suffers from an IR divergence in its first derivative.
More concretely, the one-loop effective potential, expanded at small $G$ contains the terms
\be
\delta V = -\frac{\kappa}{4}G\left[2\xi_W W
\log\left(\frac{\xi_W G W}{\mu^4}\right)+
(\xi_W W+\xi_B B)
\log\left(\frac{\xi_W G W+\xi_B G B}{\mu^4}\right)
\right]\ ,
\ee
which are responsible for producing an IR divergence in $V'$. There are no qualitative differences between this case and the Abelian one, so that again observable quantities like pole masses (for the Higgs boson and also for gauge bosons and fermions) are IR finite; IR divergences cancel out in the Nielsen identity; and the same kind of solutions discussed in Section~\ref{sec:shift} can be applied
to get rid of this complication and obtain an IR finite potential.
In particular, the field redefinition that would achieve this is now
\be
h\to h + 
\frac{\kappa}{16}h(3\xi_W g^2+\xi_B {g'}^2)
\log\left(\frac{G}{X}\right)\ ,
\label{hshift}
\ee
where $X$ is left unspecified and can be chosen at will. For instance, the choice
\be
\log X = 2(3\xi_W W+\xi_B B)\log\mu^2\ ,
\ee
modifies the one-loop anomalous dimension of the field and removes from it the $\xi_{W,Z}$ dependence, in the same way that this could be done in the Abelian model. If instead, $X$ does not 
depend explicitly on $\mu$, the anomalous dimension is not modified.

The explicit expression of the one-loop potential after shifting the field as in Eq.~(\ref{hshift}) is
\bea
V &=& -\frac12 m^2 + \frac14 \lambda h^4\nn\\
&+&\frac{\kappa}{4}\left[-12T^2(L_T-3/2)+6W^2(L_W-5/6)+3Z^2(L_Z-5/6)+H^2(L_H-3/2)\right]\nn\\
&+&\frac{\kappa}{8}G\left[2G\left(\log\frac{\xi_W G W}{\mu^4}-3\right)+G\left(\log\frac{G(\xi_W W+\xi_BB)}{\mu^4}-3\right)\right.\nn\\
&-&4\xi_W W\left(\log\frac{\xi_WXW}{\mu^4}-3\right)
-2(\xi_W W+\xi_BB)\left(\log\frac{X(\xi_WW+\xi_BB)}{\mu^4}-3\right)\nn\\
&+&\left.
2(G_{A_+}-G_{A_-})\log\frac{G_{A_+}}{G_{A_-}}+
(G_{B_+}-G_{B_-})\log\frac{G_{B_+}}{G_{B_-}}
\right]\ .
\eea

Concerning the gauge dependence of the potential, it is similarly described by Nielsen identities of the form  (\ref{NI}), one for each $\xi$ parameter (with a different $C$ function each) and the results discussed for the Abelian model carry over in a straightforward manner to the SM.

\section{Conclusions\label{sec:concl}}

In some common gauges, like Landau or Fermi gauge, Goldstone bosons are massless in the potential minimum in the broken phase
and this causes IR divergences in the calculation of the perturbative effective potential. As demonstrated recently, these divergences are spurious and can be eliminated by a simple resummation of Goldstone self-energy diagrams that otherwise lead to the breakdown of perturbation theory~\cite{GK,Martin,EEK}. 

On the other hand, when one is dealing with a potential whose minimum is generated radiatively (and this includes not only the 
well known Coleman-Weinberg model but also the SM potential at high field values) the value of the potential at that minimum (a gauge-independent quantity in principle) has a residual gauge-dependence that also needs resummation of a tower of diagrams involving Goldstone bosons. In this paper we have shown that the
resummation of IR divergences mentioned previously automatically takes care of the
residual gauge-dependence in radiative minima.

We have shown this explicitly in the case of the Abelian Higgs model in Fermi gauge, and in doing this we encountered a different IR problem: the first derivative of the  potential (and therefore also the location of the minimum) is IR divergent. This divergence can be traced back to a pole of order $p^4$ in the mixed propagator of the Goldstone bosons and longitudinal gauge bosons. As we showed, this pole is not an artifact of perturbation theory but a property of the full propagator. This mixed propagator is a specific feature of  Fermi gauge and so this IR divergence is not present in Landau gauge ($\xi \to 0$) nor in the background $R_\xi$ gauges.

Although naively this seems to be a serious pathology of Fermi gauge, interestingly the IR divergence does not propagate to physical observables. We showed explicitly that all IR divergences cancel in the physical Higgs boson mass relation as well as in the Nielsen identity, which indicates that one can extract physical 
information from the effective potential in Fermi gauge.
Still, working with an effective potential that has no well defined vacuum seems odd. Our proposal to solve this issue (besides using an IR regulator to be removed at the end of the calculations) is to remove the IR divergence of the potential by an appropriate rescaling of the Higgs field, as described in Section~\ref{sec:shift}. 
Several options for this rescaling are possible (with different advantages depending on the objective one has) and using any of them it is possible to have a well-behaved (IR finite) effective potential in Fermi gauge. It could be argued that our field redefinitions either fix the gauge (if all $\xi$ dependence of the potential is removed by the redefinition) or amount to changing to a different gauge-fixing (some kind of interpolation between Fermi and Landau gauges). In the latter case we get the best of both worlds: we inherit the good IR properties of Landau gauge and we still have a free $\xi$ parameter
to check gauge independence explicitly.

\section*{Acknowledgments}

We thank Enrico Bertuzzo, Joan Elias-Mir\'o, Gian Giudice,  Luca di Luzio, Antonio Riotto, Nikos Tetradis and Alessandro Strumia for very useful discussions. J.R.E. thanks the CERN TH-Division for hospitality and partial financial support during several stages of this project. This work has been partly supported by the ERC
grant 669668 -- NEO-NAT -- ERC-AdG-2014, the Spanish Ministry MINECO under grants  FPA2013-44773-P and
FPA2014-55613-P, the Severo Ochoa excellence program of MINECO (grant SEV-2012-0234) and by the Generalitat grant 2014-SGR-1450. MG and TK acknowledge partial support by the Munich Institute for Astro- and Particle Physics (MIAPP)
of the DFG cluster of excellence  ``Origin and Structure of the Universe''. We also acknowledge support by the German Science Foundation (DFG) within the Collaborative Research Center (SFB) 676 ‘Particles, Strings and the Early Universe.’

\appendix
\numberwithin{equation}{section}

\section{Non Perturbative Persistence of the IR Problem
\label{App:Ward}}

In this Appendix we show that the IR problem of Fermi gauge identified in the text persists non-perturbatively. In particular, we show that the full Goldstone propagator goes as $1/k^4$ in the broken phase. The proof is based on the Ward-Takahashi identities, which we 
review first.

\subsection{BRS and Ward-Takahashi Identities}

We shortly review the Ward-Takahashi identities in the Abelian Higgs  model in Fermi gauge. As usual, it is convenient to introduce an auxiliary field $B$ such that the gauge fixing term reads
\be
  {\cal L}_{GF} = F B + \frac{\xi}{2}B^2 \, ,
\ee
with $F=\partial_\mu B^\mu$ corresponding to Fermi gauge. Solving for the equation of motion for $B$ gives $B=-F/\xi$, and replacing this in the gauge fixing terms gives the usual expression. 

The Lagrangian term involving Faddeev-Popov ghosts reads 
\be
  {\cal L}_{FP} = -\bar c \left[ \frac{\delta F}{\delta B_\mu}\partial_\mu + \frac{\delta F}{\delta \phi}(i g \phi) +\frac{\delta F}{\delta \phi^*}(-i g \phi^*) \right]c 
  = -\bar c \Box c\ .
\ee
Under the BRS transformation the fields transform as
\be
  \phi_i \to \phi_i + \theta\delta_{BRS}\phi_i\,,
\ee
with a Grassmann parameter $\theta$ and where $\phi_i$ labels all fields (gauge, Higgs/Goldstone, auxiliary, ghost) with
\be
  \delta_{BRS} A_\mu = \partial_\mu c \, , \quad
  \delta_{BRS}\phi = -ig\phi c \, , \quad
  \delta_{BRS} c = 0\, , \quad
  \delta_{BRS}\bar c = B \, , \quad
  \delta_{BRS} B = 0\ .
\ee
It is convenient to split the Higgs field in components as $\phi=(v+h+i\chi)/\sqrt{2}$.
Their BRS transformation is then
\be
  \delta_{BRS} h = g \chi c \, ,\qquad 
\delta_{BRS} \chi = -g(v+h)c\,.
\ee
The generating functional in the presence of sources $J$ and $K$ reads
\be
  e^{iW[J,K]} \equiv \langle 0 | T e^{i\int d^dx (J_i(x)\phi_i(x) + K_i(x)\delta_{BRS}\phi_i(x)}|0\rangle \, ,
\ee
where $K$ sources the BRS transformation. 
The expectation value of the fields can then be written as
\be
  \phi_i(x) = \frac{\delta W}{\delta J_i(x)}\, ,
\ee
and the effective action is obtained via a Legendre transformation [we use the short-hand notation $J_i\phi_i=\int d^dx J_i(x)\phi_i(x)$]
\be
  \Gamma[\phi,K] = W[J[\phi,K],K] - J_i[\phi,K]\phi_i \, ,
\ee
Under the BRS transformations the energy functional $W$ behaves as 
\be
  W[J,K]\to W[J,K+\lambda J] \qquad \Rightarrow \qquad \delta_{BRS}W=J_i\frac{\delta W}{\delta K_i}\ ,
\ee
where we used that the BRS transformation is nilpotent,
namely $\delta_{BRS}(\delta_{BRS}\phi_i)=0$. 
At the same time, since the BRS transformation can be absorbed into the integration measure, 
one finds $\delta_{BRS}W = 0$.
Using 
\be
  \left.\frac{\delta W}{\delta K}\right|_{J=const} =  
  \left.\frac{\delta \Gamma}{\delta K}\right|_{\phi=const}\,,\qquad 
J_i = -\frac{\delta\Gamma}{\delta\phi_i}\ ,
\ee
one obtains
\be
\label{eq:BRSGamma}
0 = \frac{\delta \Gamma}{\delta\phi_i} \frac{\delta \Gamma}{\delta K_i} = \frac{\delta \Gamma}{\delta B_\mu} \frac{\delta \Gamma}{\delta K^\mu}
  +\frac{\delta \Gamma}{\delta h} \frac{\delta \Gamma}{\delta K^h}
  +\frac{\delta \Gamma}{\delta \chi} \frac{\delta \Gamma}{\delta K^\chi}
  +\frac{\delta \Gamma}{\delta \bar c} \frac{\delta \Gamma}{\delta K^{\bar c}}\ .
\ee
In the last expression we used already that the BRS transformation of $c$ and $B$ vanishes, so that
the effective action is independent of the corresponding sources $K^B$ and $K^c$.

\subsection{Ward Identities for the Gauge Boson Propagator}

Let us next derive the Ward identities for the gauge boson propagator.
In the following we assume that $v$ is the full vacuum expectation value, i.e. that
$\langle h\rangle=v$ is a solution of the equations of motion for vanishing
external source. Furthermore, the symmetry $\chi\to -\chi$, $B_\mu\to -B_\mu$ guarantees that
$\langle \chi\rangle=\langle B_\mu\rangle=0$ is a solution of the equations of motion.
This is equivalent to the condition that Fermi gauge is a `good gauge' in the sense
of Fukuda-Kugo \cite{FukudaKugo}. Finally, the symmetry $c\to -c, \bar c\to -\bar c$ guarantees that vanishing
ghost expectation value is always a solution to the equations of motion, i.e. it corresponds
to a configuration with vanishing external sources.

The full inverse propagator for the gauge field, the Goldstone boson,
and their mixing is given by
\be
  {\cal G}^{-1}(x,y) = -i\left(\begin{array}{cc} \frac{\delta^2\Gamma}{\delta B_\nu\delta B_\mu} & \frac{\delta^2\Gamma}{\delta B_\nu \delta \chi} \\
  \frac{\delta^2\Gamma}{\delta \chi\delta B_\mu }  & \frac{\delta^2\Gamma}{\delta \chi \delta \chi}\end{array}\right)\ ,
\ee
where $\Gamma$ is the effective action. One can obtain a WT identity for this propagator by taking derivatives of (\ref{eq:BRSGamma}) with respect to $B_\nu$ and $\chi$.
In addition, we take a derivative with respect to the ghost field $c$, and then set all expectation values to zero except for the Higgs, which
is assumed to be in the broken minimum. 
As discussed above, since Fermi gauge is a `good gauge', this
corresponds to a solution of the equations of motion, i.e. all sources vanish. 
The symmetry $\chi\to -\chi$, $B_\mu\to -B_\mu$ implies that second derivatives involving one
Higgs and one Goldstone or gauge field, vanish. Furthermore, all first derivatives of $\Gamma$ with respect to any field vanish
due to the on-shell stationarity of the effective action. Finally, using that $\delta\Gamma/\delta K_i=\delta W/\delta K_i=\langle\delta_{BRS}\phi_i\rangle$,
and using the BRS transformations as well as ghost number conservation, it follows that most contributions are zero for vanishing (ghost)
background field, except terms involving $\delta^2\Gamma/(\delta K_i\delta c)$ for $i=h,\chi,B_\mu$ and terms involving
$\delta^2\Gamma/(\delta c\delta\bar c)$.
 Writing the result in matrix form one obtains
\be
 {\cal G}^{-1} \, \left(\begin{array}{c} \frac{\delta^2\Gamma}{\delta K^\mu\delta c} \\ \frac{\delta^2\Gamma}{\delta K^\chi\delta c}\end{array}\right) 
+ (-i)\frac{\delta^2\Gamma}{\delta c\delta\bar c}\left(\begin{array}{c} \frac{\delta B}{\delta B_\nu} \\ \frac{\delta B}{\delta \chi} \end{array}\right) =0\,.
\ee
Here we have used already that $\delta\Gamma/\delta K^{\bar c}=\langle \delta_{BRS}\bar c\rangle=\langle \hat B\rangle=B$, where we have
denoted the field operator by a hat here, and assumed in the last equality that the auxiliary field is linear in the fundamental
fields, i.e. that the gauge fixing function $F$ is a linear function of the field variables.

Specifying to Fermi gauge where $B=-F/\xi=-\partial_\mu B^\mu/\xi$ and the ghost is a free field [such that its inverse propagator
is simply $\delta^2\Gamma/(\delta c\delta\bar c)=\delta^2 S/(\delta c\delta\bar c)=k^2$], one obtains in Fourier space
\be
 {\cal G}^{-1}(k) \, \left(\begin{array}{c} \frac{\delta^2\Gamma}{\delta K^\mu\delta c} \\ \frac{\delta^2\Gamma}{\delta K^\chi\delta c}\end{array}\right) 
- k^2\left(\begin{array}{c} k_\nu/\xi \\ 0 \end{array}\right) =0\,.
\ee
One has
$\delta^2\Gamma/[\delta K^\mu(y)\delta c(z)]=\delta[\langle \delta_{BRS} B^\mu(y)\rangle]/\delta c(z)=\delta [\partial^\mu c(y)]/\delta c(z)=\partial^\mu\delta(y-z)$.
Since the ghost is a free field one also gets $\delta^2\Gamma/[\delta K^\chi(y)\delta c(z)]=\delta [\langle(-g(v+\hat h(y))\hat c(y)\rangle]/\delta c(z)
=\delta [\langle(-g(v+\hat h(y))\rangle\langle\hat c(y)\rangle]/\delta c(z) =-gv\delta(y-z)$. In the last step we used $\langle h \rangle=0$, i.e. the assumption
that one expands around the position of the broken minimum. In Fourier space this gives
\be
 {\cal G}^{-1}(k) \, \left(\begin{array}{c} i k^\mu \\ gv\end{array}\right) 
+ k^2\left(\begin{array}{c} k_\nu/\xi \\ 0 \end{array}\right) =0\,.
\ee
It is easy to check that the WT identity is fulfilled at tree-level, where
\be
  {\cal G}_0^{-1}(k) = -i\left(\begin{array}{cc} -g_{\mu\nu}k^2+k_\mu k_\nu-k_\mu k_\nu/\xi+g^2v^2g_{\mu\nu} & -ik_\nu g v \\
   ik_\mu g v & k^2-m_\chi^2\end{array}\right)\ ,
\ee
and $m_\chi^2=0$ in the broken phase.

To see how the $k^4$ pole arises in general, 
we can decompose the propagator into transverse and 
longitudinal parts (using the short-hand notation $\hat k_\mu\equiv k_\mu/\sqrt{k^2}\equiv k_\mu/k$) as
\be
  {\cal G}(k) =  
\left(\begin{array}{cc} g_{\mu\nu}-\hat k_\mu \hat k_\nu& 0 \\ 0 & 0\end{array}\right)
{\cal G}_T(k) 
  + \left(\begin{array}{cc} -i\hat k_\mu & 0 \\ 0 & 1\end{array}\right)
\,{\cal G}_L(k)\,\left(\begin{array}{cc} i \hat k_\nu & 0 \\ 0 & 1\end{array}\right) \, ,
\ee
where ${\cal G}_T(k)$ is a function but ${\cal G}_L(k)$ still has a $2\times2$ matrix structure
and both are Lorentz scalars. 
The above WT identity then takes the form
\be
 {\cal G}_L^{-1}(k) \, \left(\begin{array}{c} k \\ gv\end{array}\right) 
-i\,\frac{k^3}{\xi} \left(\begin{array}{c} 1  \\ 0 \end{array}\right) =0\,.
\ee
Using the fact that the Goldstone mass vanishes in the broken phase, the longitudinal 
propagator must have the form 
\bea
\label{eq:TW_final}
  {\cal G}_{L}^{-1}(k) &=& \frac{i}{\xi}k^2
\left(\begin{array}{cc} 1 & 0 \\ 0 & 0 \end{array}\right) - i A(k)
\left(\begin{array}{cc} g^2v^2 & -k  g v \\ -k g v & k^2\end{array}\right)
,
\label{NPprop}
\eea
with some function $A(k)$ that has no poles at $k^2=0$.
This form means that ${\cal G}_L$ has a pole that goes (at least) like $1/k^4$, as $\det {\cal G}_L=\xi/[k^4 A(k)]$.

Another way of understanding Eq.~(\ref{NPprop}) is in terms of operators. The WT equation states that 
the operator $| D_\mu \phi |^2$ that induces the second contribution in (\ref{eq:TW_final}) is radiatively corrected while the gauge breaking term is not. The running of the parameter $\xi$ hence stems from  wave-function renormalization
of the gauge fields.
We explicitly confirmed this at the one-loop level.

\section{Higgs Self-energy and Correction to the Kinetic Term}\label{App:Higgs}

In this Appendix we present the full self-energy $\Sigma$ of the physical Higgs at one-loop in the Abelian Higgs model. We separate $\Sigma$ in different pieces as
\be
  \Sigma = \Sigma_{LL}+2\Sigma_{LT}+\Sigma_{TT}+\Sigma_{HH}+\Sigma_{L}+\Sigma_{T}+\Sigma_{H}\ ,
\ee
where the indices denote the type of fields propagating in the loop (one index for loops with a quartic vertex and two indices for loops with cubic vertices). Here $L$ labels the mixed longitudinal and Goldstone fields, $T$ the transverse part of the gauge field, and $H$ the physical Higgs.
We find [with the squared masses $H$, $G$ and $B$ as defined in (\ref{masses})]
\bea
  \Sigma_{HH}(p^2) &=& i\mu^{4-d}\, 18\lambda^2h^2\int \frac{d^dk}{(2\pi)^d}\frac{1}{(k_-^2-H)(k_+^2-H)}\ , \nn\\
  \Sigma_{TT}(p^2) &=& i\mu^{4-d}\,\frac{2B^2}{h^2}\int \frac{d^dk}{(2\pi)^d}\frac{1}{(k_-^2-B)(k_+^2-B)}\left[d-2+\frac{(k_+^2+k_-^2-p^2)^2}{4k_-^2k_+^2}\right]\ , \nn\\
  \Sigma_{LT}(p^2) &=& i\mu^{4-d}\,\frac{B}{2h^2}\int \frac{d^dk}{(2\pi)^d}\frac{p^2(p^2-4k_-^2)(k_+^4+B G \xi)}{k_-^2k_+^2(k_-^2-B)(k_+^4-Gk_+^2+B G \xi)}\ , \\
  \Sigma_{LL}(p^2) &=& i\mu^{4-d}\int \frac{d^dk}{(2\pi)^d}\frac{1}{(k_-^4-Gk_-^2+B G \xi)(k_+^4-Gk_+^2+B G \xi)}\sum_{i,j=1}^5 E_{ij} k_-^{2i-4}k_+^{2j-4}\ , \nn
\eea
where the $E_{ij}$ entering $\Sigma_{LL}(p^2)$ is given by the matrix
\be
E \equiv \left(\begin{array}{ccccc} 
 B^2 G^2 p^4 \xi ^2 & -2 B^2 G^2 p^2 \xi ^2 & B G \xi  \left(p^4+B G \xi \right) & -2 B G p^2 \xi  & B G \xi \\
 -2 B^2 G^2 p^2\xi ^2 & B^2 \left(2 G^2+\Delta ^4\right) \xi ^2&-B \xi  \left(\Delta ^4+B G \xi \right)&2 B \Delta ^2 \xi &-B \xi \\
 B G \xi  \left(p^4+BG \xi \right)&-B \xi  \left(\Delta ^4+B G \xi \right)&4 \lambda^2 h^4&0&0\\
-2 B G p^2 \xi &2 B \Delta ^2 \xi &0&0&0\\
B G \xi &-B \xi &0&0&0 \end{array}\right)\ ,
\ee
with $\Delta^2\equiv p^2-2\lambda h^2$, $k_\pm^2\equiv (k \pm p/2)^2$,
and
\bea
\Sigma_H &=& 3i\lambda\mu^{4-d}\int\frac{d^dk}{(2\pi)^d}\frac{1}{k^2-H}\ ,  \nn\\
\Sigma_T &=& i\frac{(d-1)B}{h^2}\mu^{4-d}\int\frac{d^dk}{(2\pi)^d}\frac{1}{k^2-B}\ , \nn\\
\Sigma_L &=& i\mu^{4-d}\int \frac{d^dk}{(2\pi)^d}\frac{k^2(\lambda h^2+\xi B)-\xi B(\lambda h^2+G)}{h^2(k^4-Gk^2+B G \xi)}\ .
\eea
All loop integrals can be reduced to the elementary one-loop functions
\bea
  A(X) &\equiv& -i \mu^{4-d} 16\pi^2 \int \frac{d^dk}{(2\pi)^d} \, \frac{1}{k^2-X} 
= m^2\left[\Delta_\epsilon + 1 - \log(X/\mu^2)\right] ,\\ 
  B(X_1,X_2,p^2) &\equiv& -i \mu^{4-d} 16\pi^2 \int \frac{d^dk}{(2\pi)^d} \, \frac{1}{(k_-^2-X_1-i\varepsilon)}\, \frac{1}{(k_+^2-X_2-i\varepsilon)} \nn\\
  &=& \Delta_\epsilon - \int_0^1 dx \log\left[\frac{X_1(1-x)+X_2x-x(1-x)p^2-i\varepsilon}{\mu^2}\right] ,
\eea
with $\Delta_\epsilon \equiv 2/(4-d) -\gamma_E + \log 4\pi$. After factorizing 
\be
N(k)\equiv k^4-Gk^2+B G \xi = (k^2-G_-^2)(k^2-G_+^2) ,
\ee 
with $G_\pm$ as defined in (\ref{Gpm}), the reduction can be done in several ways. 
For example, 
in $\Sigma_{TT}$ one can write the numerator of the last term as
\be
  (k_+^2+k_-^2-p^2)^2=2k_+^2k_-^2+p^4+k_+^2(k_+^2-B)+k_-^2(k_-^2-B)+(B-2p^2)(k_+^2+k_-^2)\,,
\ee
and then split the integral into contributions from each summand. Similarly, in $\Sigma_{LT}$ one can add and subtract $Gk_+^2$ in the right-most bracket of the numerator, and in
$\Sigma_L$ one can write $k^2=[(k^2-G_-)+(k^2-G_+)]/2+G/2$. One then finds
\bea
 \Sigma_{HH}(p^2) &=&- 18\kappa\lambda^2h^2 B(H,H,p^2) , \nn\\
 \Sigma_{TT}(p^2) &=& -\frac{2\kappa B^2}{v^2}\bigg[B(T,T,p^2)(d-3/2)+\frac{(B-2p^2)}{2}B^{2,1}(0,B,B,p^2) \nn\\
 && {} +\frac12 B(0,B,0) +\frac{p^4}{4}B^{2,2}(0,B,0,B,p^2)\bigg],\nn\\
 \Sigma_{LT}(p^2) &=& -\kappa\frac{p^2B}{2h^2}\bigg[p^2B^{2,1}(0,B,0,p^2)-4B(0,B,p^2) \nn \\
 && {} +Gp^2B^{2,2}(0,B,G_+,G_-,p^2)  
 -4GB^{2,1}(G_+,G_-,B,p^2)\bigg] ,
\eea
and
\bea
\Sigma_H &=& -3\kappa\lambda H A(H), \nn\\
\Sigma_T &=& -\kappa\frac{B^2}{h^2}(d-1)A(B),\nn\\
\Sigma_L &=& -\kappa\frac{(G \lambda h^2 -\xi B G -2\xi B \lambda h^2)}{2h^2}B(G_+,G_-,0) -\frac{\kappa}{2h^2}[A(G_+)+A(G_+)] ,
\eea
where $\kappa=1/(16\pi^2)$ and
\bea
  B^{2,1}(X_1,X_2,X_3,p^2) &\equiv& -i \mu^{4-d} 16\pi^2 \int \frac{d^dk}{(2\pi)^d} \, \frac{1}{(k_-^2-X_1-i\varepsilon)}\frac{1}{(k_-^2-X_2-i\varepsilon)}\, \frac{1}{(k_+^2-X_3-i\varepsilon)} \nn\\
 &=& \frac{1}{X_1-X_2}[B(X_1,X_3,p^2)-B(X_2,X_3,p^2)]\;.
\eea
Similarly $B^{2,2}$ contains two propagators involving $k_-^2$ and $k_+^2$, respectively, and is
\be
  B^{2,2}(X_1,X_2,X_3,X_4,p^2) = \frac{B(X_1,X_3,p^2)-B(X_1,X_4,p^2)-B(X_2,X_3,p^2)+B(X_2,X_4,p^2)}{(X_1-X_2)(X_3-X_4)} .
\ee
The piece $\Sigma_{LL}$ can be reduced by first rewriting the powers of  $k_\pm^2$ for $i=4,5$ or $j=4,5$ as
\bea
  k_\pm^4 &=& N(k_\pm) +Gk_\pm^2-BG\xi , \nn\\
  k_\pm^6 &=& (k_\pm^2+G)N(k_\pm)+G(G-B\xi)k_\pm^2-B G^2\xi \;.
\eea
Then one obtains integrals that can be reduced as above, for example
\bea
  -i \mu^{4-d} 16\pi^2 \int \frac{d^dk}{(2\pi)^d} \, \frac{k_-^2}{N(k_+)N(k_-)}&=&\frac12\Big[B^{2,1}(G_+,G_-,G_+,p^2)+B^{2,1}(G_+,G_-,G_-,p^2)\nn\\
  && +G\ B^{2,2}(G_+,G_-,G_+,G_-,p^2)\Big]\ ,
\eea
where we rewrote $k_-^2$ in the numerator as done for $\Sigma_L$. Another useful relation is
\be
  \int \frac{d^dk}{(2\pi)^d} \, \frac{k_-^2}{N(k_+)} = \int \frac{d^dk}{(2\pi)^d} \, \frac{k^2+p^2}{N(k)} \ .
\ee
The full result for $\Sigma_{LL}$ is straightforward but too lengthy to report.
A useful check is that $\Sigma(p^2=0)=V''_{1}$. To get the self-energy at finite $p^2$ in the
limit $G\to 0$ we used the expansion
\bea
  \lefteqn{\int_0^1 dx \log[a(1-x)+bx-x(1-x)c] =} \nn\\
  &=& -2 + \frac12\log(ab) + \frac{a-b}{2c}\log\left(\frac{a}{b}\right) + \frac{S}{2c}\log\left(\frac{c-a-b+S}{c-a-b-S}\right) \nn\\
  &= & -2 + \log(-c) + \frac{1}{c}\left[a(\log a-1)+b(\log b-1) -(a+b)\log(-c)\right] \nn\\
  && - \frac{1}{2c^2}\left\{a^2+b^2+2ab\left[\log(ab)-2\log(-c)\right]\right\} + {\cal O}\left(\frac{1}{c^3}\log c\right)\ ,
\eea
for small $a, b$, where $S=\sqrt{(c-a-b)^2-4ab}$.

From the self-energies one can get the corrections to the kinetic term in the effective action
\be
  Z(h) = 1  - \frac{d\Sigma(p^2)}{dp^2}\Big|_{p^2=0}\,.
\ee
We find
\be
  Z = 1 + Z_{LL}+2Z_{LT}+Z_{TT}+Z_{HH} \,,
\ee
with the explicit results (for the Abelian Higgs model) 
\bea\label{eq:ZLLU1}
  Z_{LL} &=& \kappa\frac{\xi B}{h^2}\Delta_\epsilon +\frac{\kappa}{3\pi^2h^2} \Bigg\{G
  - 3 \xi B \left[\frac{G_+(L_{G_+}+1)-G_-(L_{G_-}+1)}{\Delta G_\pm}\right] \nn\\
  && {}  + \frac{m^2}{\Delta G_\pm^4} \Bigg[ 
  \left(2 \Delta G_\pm^2+m^2 G\right)(G-6\xi B)(G-\xi B)
  \nn\\
 && {}
  +6 \xi^2 B^2 G \left(\Delta G_\pm^2 + 2 m^2 \xi B \right)\frac{(L_{G_+}-L_{G_-})}{\Delta G_\pm}  \Bigg] \Bigg\}\,,
\\
 2Z_{LT} &=& - \frac{3\kappa B}{h^2}\Delta_\epsilon + \frac{3\kappa B}{2h^2B_\xi}\Bigg[2(B+\xi G)\left(L_B-\frac56\right)
   \nn\\
 && -G\left(L_{G_+}+L_{G_-}-\frac53\right)-G^2(1-2\xi)\left(\frac{L_{G_+}-L_{G_-}}{\Delta G_\pm}\right)\Bigg]
 \,,
\eea
and 
\be
  Z_{TT} =  \frac{5\kappa B}{2h^2} \ , \quad\quad\quad\quad
  Z_{HH} = \frac{3\kappa \lambda^2h^2}{ H} \,,
\ee
where $L_{G_\pm}\equiv \log(G_\pm/\mu^2)$, $L_B\equiv \log(B/\mu^2)$, $B_\xi\equiv B+G(\xi-1)$ and $\Delta G_\pm\equiv G_+-G_-$.
For large field values one can approximate $G=-m^2+\lambda h^2\to \lambda h^2$ and neglect the second and third 
line in (\ref{eq:ZLLU1}) which are proportional to the quadratic mass parameter $m^2$. In that limit, $Z_{HH}\ra \lambda\kappa$.

For completeness, we also present the off-shell Higgs self-energy for $h\to v$
\bea
 \Sigma(p^2)|_{h\to v} &=& \frac{\kappa v^2}{2}\Bigg\{2\left[-26\lambda^2+5\lambda g^2\xi +g^2(3-\xi)P^2-9g^4\right]\frac{1}{\epsilon}+2P^4 \nn\\
 && {} +\left[\log(p^2/\mu^2)+i\pi\right](4\lambda^2-P^4) +\xi g^2(P^2-5\lambda)\log\left(\xi G B/\mu^4\right)\nn\\
 && {} + \lambda^2 \left[-20 + 12L_H + 36 \int_0^1 dx \log\left[(H-x(1-x)p^2)/\mu^2\right] \right]\nn\\
 && {} +6g^4(1+L_B)+2g^2P^2(1-\xi-L_B)+10\lambda g^2\xi 
 \nn\\
 && {} + (P^4-4P^2 g^2+12g^4) \int_0^1 dx \log\left[(B-x(1-x)p^2)/\mu^2\right]\nn\\
 && {} - \xi g^2 \frac{(P^2-2\lambda )^2}{P^2} \bigg[ \left(1-\frac{g^2\xi}{P^2}\right)\left[ \log\left(\xi G B/p^4\right) 
- 2i\pi  \right]-2 \bigg] \Bigg\}\ ,
\eea
where $P^2\equiv p^2/v^2$.

\section{Nielsen Identity for the Kinetic Term}\label{app:NielsenKin}

The Nielsen identity for the kinetic term $\delta {\cal L}_K=Z(h) (\partial_\mu h)^2/2$ reads \cite{GK,Metaxas}
\be
  \xi \frac{\partial Z}{\partial\xi} =  - C \frac{\partial Z}{\partial h} -2Z\frac{\partial C}{\partial h} + D \frac{\partial V}{\partial h}
 + \widetilde D \frac{\partial^2 V}{\partial h^2}\ ,
\ee
with coefficients given by a gradient expansion of (\ref{eq:NielsenC}),
\be
  C \to C + D (\partial_\mu h)^2 - \partial_\mu \left( \widetilde D \partial^\mu h\right) + {\cal O}(\partial^4)\;.
\ee 
Note that Ref.~\cite{Metaxas} did not include the total derivative term above, which is relevant as it is required to describe the full $\xi$ dependence of the function $Z$.

At one-loop, the contribution at zeroth order in gradients is given by
the one-loop expressions of $C,D$ and $\widetilde D$. The $C_1$ function is given in Eq.~\eqref{eq:NielsenC1}, and $D_1,\widetilde D_1$ are 
\bea
  \widetilde D_{1} &=& -\frac{\kappa G \xi }{3}\Bigg[
  \frac{2g^2}{G^2}+
\frac{2(\xi B-G)B}{\Delta G_\pm^4} (\lambda -4 g^2 \xi )         
-3 B \xi \left[g^2 G^2 + 2\lambda B ( 2 \xi B -G )\right] \frac{\log(G_+/G_-)}{\Delta G_\pm^5}  \Bigg] ,\nn\\
  D_{1} &=& \frac{\kappa\xi G}{6 g^2 h B_\xi^2 
 }\Bigg\{ 9 g^4  \log\left(\frac{G\xi}{B}\right) 
   + \frac{2B_\xi}{ \Delta G_\pm^6} \Bigg[-\lambda^2 B^2 B_\xi(3G^2-4G B\xi+28B^2\xi^2)  \nn\\
  && {}  + 
   g^4 G^2 \Big[3G^3(-4+\xi)+212 B^3\xi^2+G^2B(21+80\xi-8\xi^2)-4GB^2\xi(38+17\xi+19\xi^2) 
   \Big] \nn\\
  &&    -    8 g^2 \lambda\xi B^2 B_\xi (G-B \xi)(G-24B\xi)
\Bigg]  
\nn\\
&&- 3 \frac{\log(G_+/G_-)}{\Delta G_\pm^7}\Bigg[
      g^4 \Big[
      \frac{3G}{2}(2\xi-1)\Delta G_\pm^4\left[3\Delta G_\pm^2-G^2(-1+2\xi)^2\right]
          \nn\\
  && {}   +\frac{B_\xi^2}{4G}\left[-\Delta G_\pm^6+G^2\Delta G_\pm^4(13+48\xi)+17G^4 \Delta G_\pm^2-5G^6\right]
  \Big]\nn\\
  &&{}
   -4\xi\lambda B^2 B_\xi^2 \Big[ 2\lambda B (G^2 - 3 G B \xi + 
      6 B^2 \xi^2) + g^2  G  (G^2 - 16 G B \xi +   8 B^2 \xi^2) \Big] 
  \Bigg]  \Bigg\} \ ,
\eea
where $B_\xi\equiv B + G (\xi-1)$ and $\Delta G_\pm\equiv G_+-G_-$. The small-$G$ expansion of these functions read:
\bea
\widetilde D_1 &=&  -\frac{\kappa \pi \lambda g h \xi^{1/2}}{8 G^{3/2}}-\frac{\kappa(\lambda+3g^2\xi)}{6G}-\frac{\kappa \lambda\pi}{64 (BG\xi)^{1/2}}+{\cal O}(G^0)\ ,\nn\\
D_1 &=& \frac{3\kappa\pi \lambda^2g h^2\xi^{1/2}}{16 G^{5/2}}+\frac{\kappa\lambda h(\lambda+3g^2\xi)}{3 G^2}
        +\frac{\kappa\pi\lambda(9\lambda+16g^2\xi)}{128 g \xi^{1/2} G^{3/2}} +{\cal O}(G^{-1})\;.
\eea
With these expressions one can check explicitly that the Nielsen identity for $Z$ is fulfilled perturbatively at one-loop.

\section{Kinetic Term for the SM in Fermi Gauge}

In the Standard Model, the kinetic term is readily obtained from the corresponding expressions in the
Abelian Higgs model, given in Appendix~\ref{App:Higgs}.
 We find
\be
  Z = 1 + Z_{LL}+2Z_{L^+L^-}+2Z_{LT}+4Z_{L^+T^-}+Z_{TT}+2Z_{T^+T^-}+Z_{HH}+Z_{tt} \,,
\ee
with
\be
\begin{array}{rclrcl}
  Z_{HH} &=& Z_{HH}^{U(1)}\,,\quad
  &
  Z_{TT} &=& Z_{TT}^{U(1)}\Big|_{B\to Z} \,,
  \nn\\
  Z_{T^+T^-} &=& Z_{TT}^{U(1)}\Big|_{B\to W} \,,
  & Z_{LT} &=& Z_{LT}^{U(1)}\Big|_{B\to Z, \xi\to \xi_{\rm eff}}  \,,
  \nn\\
  Z_{L^+T^-} &=& Z_{LT}^{U(1)}\Big|_{B\to W, \xi\to \xi_W} \,,
 & Z_{LL} &=& Z_{LL}^{U(1)}\Big|_{B\to Z, \xi\to \xi_{\rm eff}}   \,,
  \nn\\
  Z_{L^+L^-} &=& Z_{LL}^{U(1)}\Big|_{B\to W, \xi\to \xi_W} \,,\quad\quad
  & Z_{tt} &=& \frac{6\kappa T}{h^2}\Delta_\epsilon - \frac{6\kappa T}{h^2}\left[1/4 + \log(T/\mu^2)\right] \ ,
\end{array}
\ee
where $W=g^2h^2/4$, $Z=(g^2+{g'}^2)h^2/4$, $B=Z-W$, $T=y_t^2h^2/2$ and $\xi_{\rm eff}\equiv (\xi_B B + \xi_W W)/(B+W)=\xi_B s_W^2+\xi_W c_W^2$.
The divergent part is given by (note that the TT and hh parts are UV finite)
\bea
 Z^{\rm div} &=& \frac{\kappa}{h^2}\left(\xi_{\rm eff} Z + 2\xi_W W - 3Z - 6W + 6 T\right)\Delta_\epsilon \nn\\
 &=& \frac{\kappa}{4}\left(\xi_B{g'}^2 + 3\xi_W g^2 - 3{g'}^2 - 9g^2 + 12y_t^2\right)\Delta_\epsilon \,.
\eea
This is consistent with Eq.~(2.47) of \cite{DiLuzio}. In particular, the field renormalization $Z_h$ cancels the divergences in $Z$.



\begin{thebibliography}{10}


\bibitem{higgsD}
  G.~Aad {\it et al.}  [ATLAS Collaboration],
  Phys.\ Lett.\ B {\bf 716} (2012) 1
  \arXiv{1207.7214}{hep-ex};
  S.~Chatrchyan {\it et al.}  [CMS Collaboration],
  Phys.\ Lett.\ B {\bf 716} (2012) 30
  \arXiv{1207.7235}{hep-ex}.
  
\bibitem{mhLHC}
  G.~Aad {\it et al.} [ATLAS and CMS Collaborations],
  Phys.\ Rev.\ Lett.\  {\bf 114} (2015) 191803
  \arXiv{1503.07589}{hep-ex}.
  
\bibitem{mtcomb}
  [ATLAS and CDF and CMS and D0 Collaborations],
  \arXiv{1403.4427}{hep-ex}.
    
\bibitem{stab0}
  G.~Degrassi, S.~Di Vita, J.~Elias-Mir\'o, J.~R.~Espinosa, G.~F.~Giudice, G.~Isidori and A.~Strumia,
  JHEP {\bf 1208} (2012) 098
  \arXiv{1205.6497}{hep-ph}.


\bibitem{stab1}
  D.~Buttazzo, G.~Degrassi, P.~P.~Giardino, G.~F.~Giudice, F.~Sala, A.~Salvio and A.~Strumia,
  JHEP {\bf 1312} (2013) 089
 \arXiv{1307.3536}{hep-ph}.

\bibitem{stab2}
  F.~Bezrukov, M.~Y.~Kalmykov, B.~A.~Kniehl and M.~Shaposhnikov,
  JHEP {\bf 1210} (2012) 140
  \arXiv{1205.2893}{hep-ph}.

\bibitem{stab3}
  M.~Holthausen, K.~S.~Lim and M.~Lindner,
  JHEP {\bf 1202} (2012) 037
  \arXiv{1112.2415}{hep-ph};
  J.~Elias-Mir\'o, J.~R.~Espinosa, G.~F.~Giudice, G.~Isidori, A.~Riotto and A.~Strumia,
  Phys.\ Lett.\ B {\bf 709} (2012) 222
  \arXiv{1112.3022}{hep-ph};
  S.~Alekhin, A.~Djouadi and S.~Moch,
  Phys.\ Lett.\ B {\bf 716} (2012) 214
  \arXiv{1207.0980}{hep-ph}.

\bibitem{stab4}
  A.~V.~Bednyakov, B.~A.~Kniehl, A.~F.~Pikelner and O.~L.~Veretin,
  Phys.\ Rev.\ Lett.\  {\bf 115} (2015) 20,  201802
  \arXiv{1507.08833}{hep-ph}.

\bibitem{stab5}
  L.~Di Luzio, G.~Isidori and G.~Ridolfi,
  Phys.\ Lett.\ B {\bf 753} (2016) 150
  \arXiv{1509.05028}{hep-ph}.
\bibitem{cosmostab}
  J.~R.~Espinosa, G.~F.~Giudice and A.~Riotto,
  JCAP {\bf 0805} (2008) 002
  \arXiv{0710.2484}{hep-ph};
  O.~Lebedev and A.~Westphal,
  Phys.\ Lett.\ B {\bf 719} (2013) 415
  \arXiv{1210.6987}{hep-ph};
  A.~Kobakhidze and A.~Spencer-Smith,
  Phys.\ Lett.\ B {\bf 722} (2013) 130
  \arXiv{1301.2846}{hep-ph};
  M.~Fairbairn and R.~Hogan,
  Phys.\ Rev.\ Lett.\  {\bf 112} (2014) 201801
  \arXiv{1403.6786}{hep-ph};
  K.~Enqvist, T.~Meriniemi and S.~Nurmi,
  JCAP {\bf 1407} (2014) 025
  \arXiv{1404.3699}{hep-ph};
  A.~Hook, J.~Kearney, B.~Shakya and K.~M.~Zurek,
  JHEP {\bf 1501} (2015) 061
  \arXiv{1404.5953}{hep-ph};
  M.~Herranen, T.~Markkanen, S.~Nurmi and A.~Rajantie,
  Phys.\ Rev.\ Lett.\  {\bf 113} (2014) 21,  211102
  \arXiv{1407.3141}{hep-ph};
  F.~Bezrukov, J.~Rubio and M.~Shaposhnikov,
  Phys.\ Rev.\ D {\bf 92} (2015) 8,  083512
  \arXiv{1412.3811}{hep-ph};
  A.~Shkerin and S.~Sibiryakov,
  Phys.\ Lett.\ B {\bf 746} (2015) 257
  \arXiv{1503.02586}{hep-ph};
  J.~Kearney, H.~Yoo and K.~M.~Zurek,
  Phys.\ Rev.\ D {\bf 91} (2015) 12,  123537
  \arXiv{1503.05193}{hep-th};
  P.~Burda, R.~Gregory and I.~Moss,
  JHEP {\bf 1508} (2015) 114
  \arXiv{1503.07331}{hep-th};
  C.~Gross, O.~Lebedev and M.~Zatta,
  Phys.\ Lett.\ B {\bf 753} (2016) 178
  \arXiv{1506.05106}{hep-ph};
  M.~Herranen, T.~Markkanen, S.~Nurmi and A.~Rajantie,
  Phys.\ Rev.\ Lett.\  {\bf 115} (2015) 241301
  \arXiv{1506.04065}{hep-ph};
  L.~Delle Rose, C.~Marzo and A.~Urbano,
  \arXiv{1507.06912}{hep-ph};
  P.~Burda, R.~Gregory and I.~Moss,
  \arXiv{1601.02152}{hep-th};
  Y.~Ema, K.~Mukaida and K.~Nakayama,
  \arXiv{1602.00483}{hep-ph};
  K.~Kohri and H.~Matsui,
  \arXiv{1602.02100}{hep-ph}.


\bibitem{deep}
  C.~D.~Froggatt and H.~B.~Nielsen,
  Phys.\ Lett.\ B {\bf 368} (1996) 96
  \arXivold{hep-ph/9511371};
  M.~Shaposhnikov and C.~Wetterich,
  Phys.\ Lett.\ B {\bf 683} (2010) 196
  \arXiv{0912.0208}{hep-th};
  I.~Bars, P.~J.~Steinhardt and N.~Turok,
  Phys.\ Lett.\ B {\bf 726} (2013) 50
  \arXiv{1307.8106}{gr-qc};
  A.~Hebecker, A.~K.~Knochel and T.~Weigand,
  JHEP {\bf 1206} (2012) 093
  \arXiv{1204.2551}{hep-th};
  L.~E.~Iba\~nez, F.~Marchesano, D.~Regalado and I.~Valenzuela,
  JHEP {\bf 1207} (2012) 195
  \arXiv{1206.2655}{hep-ph};
  L.~E.~Iba\~nez and I.~Valenzuela,
  JHEP {\bf 1305} (2013) 064
  \arXiv{1301.5167}{hep-ph};
  A.~Hebecker, A.~K.~Knochel and T.~Weigand,
  Nucl.\ Phys.\ B {\bf 874} (2013) 1
  \arXiv{1304.2767}{hep-th};
  M.~Ibe, S.~Matsumoto and T.~T.~Yanagida,
  Phys.\ Lett.\ B {\bf 732} (2014) 214
  \arXiv{1312.7108}{hep-ph};
  F.~D'Eramo, L.~J.~Hall and D.~Pappadopulo,
  JHEP {\bf 1506} (2015) 117
  \arXiv{1502.06963}{hep-ph};
  J.~R.~Espinosa, J.~F.~Fortin and M.~Trepanier,
  \arXiv{1508.05343}{hep-th}.
  
  
\bibitem{cosmostab2}
  J.~R.~Espinosa, G.~F.~Giudice, E.~Morgante, A.~Riotto, L.~Senatore, A.~Strumia and N.~Tetradis,
  JHEP {\bf 1509} (2015) 174
  \arXiv{1505.04825}{hep-ph}.


\bibitem{Jackiw}
  R.~Jackiw,
  Phys.\ Rev.\ D {\bf 9} (1974) 1686.
  
\bibitem{LitVgauge}
  D.~Johnston,
  Nucl.\ Phys.\ B {\bf 253} (1985) 687;
  G.~Thompson and H.~L.~Yu,
  Phys.\ Rev.\ D {\bf 31} (1985) 2141;
  G.~Kunstatter and H.~P.~Leivo,
  Phys.\ Lett.\ B {\bf 183} (1987) 75;
  J.~R.~S.~Do Nascimento and D.~Bazeia,
  Phys.\ Rev.\ D {\bf 35} (1987) 2490;
  D.~Boyanovsky, D.~Brahm, R.~Holman and D.~S.~Lee,
  Phys.\ Rev.\ D {\bf 54} (1996) 1763
  \arXivold{hep-ph/9603337};
  C.~Contreras and L.~Vergara,
  Phys.\ Rev.\ D {\bf 55} (1997) 5241
   Erratum: [Phys.\ Rev.\ D {\bf 56} (1997) 6714]
  \arXivold{hep-th/9610109};
  D.~Boyanovsky, W.~Loinaz and R.~S.~Willey,
  Phys.\ Rev.\ D {\bf 57} (1998) 100
  \arXivold{hep-ph/9705340};
  R.~Haussling and S.~Kappel,
  Eur.\ Phys.\ J.\ C {\bf 4} (1998) 543
  \arXivold{hep-th/9707165};
  O.~M.~Del Cima, D.~H.~T.~Franco and O.~Piguet,
  Nucl.\ Phys.\ B {\bf 551} (1999) 813
  \arXivold{hep-th/9902084};
  J.~Baacke and K.~Heitmann,
  Phys.\ Rev.\ D {\bf 60} (1999) 105037
  \arXivold{hep-th/9905201};
  L.~P.~Alexander and A.~Pilaftsis,
  J.\ Phys.\ G {\bf 36} (2009) 045006
  \arXiv{0809.1580}{hep-ph}.

  
\bibitem{Metaxas} D.~Metaxas and E.~J.~Weinberg,
  Phys.\ Rev.\ D {\bf 53}, 836 (1996)
  \arXivold{hep-ph/9507381};
  D.~Metaxas,
  Phys.\ Rev.\ D {\bf 63} (2001) 085009
  \arXivold{hep-ph/0011015}.

 
\bibitem{IRS}
  G.~Isidori, G.~Ridolfi and A.~Strumia,
  Nucl.\ Phys.\ B {\bf 609} (2001) 387
  \arXivold{hep-ph/0104016}.
  
\bibitem{PRM}
  H.~H.~Patel and M.~J.~Ramsey-Musolf,
  JHEP {\bf 1107} (2011) 029
  \arXiv{1101.4665}{hep-ph}.

\bibitem{GK}
  M.~Garny and T.~Konstandin,
  JHEP {\bf 1207} (2012) 189
  \arXiv{1205.3392}{hep-ph}.
  
 
\bibitem{AFS}
  A.~Andreassen, W.~Frost and M.~D.~Schwartz,
  Phys.\ Rev.\ D {\bf 91} (2015) 1,  016009
  \arXiv{1408.0287}{hep-ph};
  Phys.\ Rev.\ Lett.\  {\bf 113} (2014) 24,  241801
  \arXiv{1408.0292}{hep-ph}.


\bibitem{Lifetime}
  A.~D.~Plascencia and C.~Tamarit,
  \arXiv{1510.07613}{hep-ph};
  Z.~Lalak, M.~Lewicki and P.~Olszewski,
  \arXiv{1605.06713}{hep-ph}.
  
  
\bibitem{CW}
  S.~R.~Coleman and E.~J.~Weinberg,
  Phys.\ Rev.\ D {\bf 7} (1973) 1888.
  
\bibitem{Martin}
  S.~P.~Martin,
  Phys.\ Rev.\ D {\bf 90} (2014) 1,  016013
  \arXiv{1406.2355}{hep-ph}.

\bibitem{EEK}
  J.~Elias-Mir\'o, J.~R.~Espinosa and T.~Konstandin,
  JHEP {\bf 1408} (2014) 034
  \arXiv{1406.2652}{hep-ph}.


\bibitem{Loinaz}
  W.~Loinaz and R.~S.~Willey,
  Phys.\ Rev.\ D {\bf 56} (1997) 7416
  \arXivold{hep-ph/9702321}.
  

\bibitem{Nielsen}
  N.~K.~Nielsen,
  Nucl.\ Phys.\ B {\bf 101} (1975) 173.
  
  
\bibitem{Aitchison}
  I.~J.~R.~Aitchison and C.~M.~Fraser,
  Annals Phys.\  {\bf 156} (1984) 1.



\bibitem{OtherIR}
  C.~Ford, D.~R.~T.~Jones, P.~W.~Stephenson and M.~B.~Einhorn,
  Nucl.\ Phys.\ B {\bf 395} (1993) 17
  \arXivold{hep-lat/9210033};
  M.~B.~Einhorn and D.~R.~T.~Jones,
  JHEP {\bf 0704} (2007) 051
  \arXivold{hep-ph/0702295}.




\bibitem{Pila}
  A.~Pilaftsis and D.~Teresi,
  Nucl.\ Phys.\ B {\bf 906} (2016) 381
  \arXiv{1511.05347}{hep-ph};
  G.~Mark\'o, U.~Reinosa and Z.~Sz\'ep,
  \arXiv{1604.04193}{hep-ph}.
\bibitem{Martin2}
  N.~Kumar and S.~P.~Martin,
  \arXiv{1605.02059}{hep-ph}.
  
\bibitem{Kim}
  C.~Kim,
  Phys.\ Rev.\ D {\bf 90} (2014) no.6,  067701
   Erratum: [Phys.\ Rev.\ D {\bf 90} (2014) no.8,  089903]
  \arXiv{1410.0453}{hep-ph}.
  


\bibitem{DJ}
  L.~Dolan and R.~Jackiw,
  Phys.\ Rev.\ D {\bf 9} (1974) 3320.
  
\bibitem{DiLuzio}
  L.~Di Luzio and L.~Mihaila,
  JHEP {\bf 1406} (2014) 079
  \arXiv{1404.7450}{hep-ph}.





\bibitem{FukudaKugo}
  R.~Fukuda and T.~Kugo,
  Phys.\ Rev.\ D {\bf 13} (1976) 3469.





\bibitem{Johnston}
  D.~Johnston,
  Phys.\ Lett.\ B {\bf 186} (1987) 185.
  
\bibitem{Dine}
  M.~Dine, P.~Draper, H.~E.~Haber and L.~S.~Haskins,
  \arXiv{1607.06995}{hep-th}.
  
\bibitem{Gambino}
  P.~Gambino and P.~A.~Grassi,
  Phys.\ Rev.\ D {\bf 62} (2000) 076002
  \arXivold{hep-ph/9907254}.
  
\bibitem{CEQR}
  J.~A.~Casas, J.~R.~Espinosa, M.~Quir\'os and A.~Riotto,
  Nucl.\ Phys.\ B {\bf 436} (1995) 3
   Erratum: [Nucl.\ Phys.\ B {\bf 439} (1995) 466]
  \arXivold{hep-ph/9407389}.
  
  \bibitem{SZ}
  A.~Sirlin and R.~Zucchini,
  Nucl.\ Phys.\ B {\bf 266} (1986) 389.
  
\bibitem{shift}
  S.~R.~Coleman, J.~Wess and B.~Zumino,
  Phys.\ Rev.\  {\bf 177} (1969) 2239;
  C.~Arzt,
  Phys.\ Lett.\ B {\bf 342} (1995) 189
  \arXivold{hep-ph/9304230}.
   
\bibitem{NielsenRedef}
  N.~K.~Nielsen,
  Phys.\ Rev.\ D {\bf 90} (2014) 3,  036008
  \arXiv{1406.0788}{hep-ph}.
  
\end{thebibliography}
\end{document}